\documentclass[10pt,journal,compsoc]{IEEEtran}
\usepackage{graphicx}
 \usepackage{float}
 \usepackage{algorithm}
\usepackage{algorithmic}
 \usepackage{subfigure}
\usepackage{amsmath,amssymb,amsfonts}
\renewcommand{\algorithmicrequire}{ \textbf{Hyperparameters:}}

 \usepackage{url}
 \usepackage{amsmath,bm}
 \usepackage{url}
\usepackage{multirow}
\ifCLASSOPTIONcompsoc
  \usepackage[nocompress]{cite}
\else
  \usepackage{cite}
\fi

\ifCLASSINFOpdf
\else
\fi

\begin{document}
%
\title{Federated Multi-Discriminator BiWGAN-GP based Collaborative Anomaly Detection for Virtualized Network Slicing}

\author{Weili~Wang,
        Chengchao~Liang,
        Lun~Tang,
        Halim~Yanikomeroglu,~\IEEEmembership{Fellow,~IEEE},
        Qianbin~Chen,~\IEEEmembership{Senior Member,~IEEE}

\thanks{W. Wang, C. Liang, L. Tang and Q. Chen are with the School of Communication and Information Engineering and the Key Laboratory of Mobile
Communication,
Chongqing University of Posts and Telecommunications, Chongqing
400065, China (e-mail: 1961797154@qq.com; liangcc@cqupt.edu.cn; tangluncq@163.com; cqb@cqupt.edu.cn).}
\thanks{H. Yanikomeroglu is with the Department of Systems and
Computer Engineering, Carleton University, Ottawa, ON, Canada (e-mail:
halim@sce.carleton.ca).}
}

%

%

\IEEEtitleabstractindextext{%
\begin{abstract}
Virtualized network slicing allows a multitude of logical networks to be created on a common substrate infrastructure to support diverse services. A virtualized network slice is a logical combination of multiple virtual network functions, which run on virtual machines (VMs) as software applications by virtualization techniques. As the performance of network slices hinges on the normal running of VMs, detecting and analyzing anomalies in VMs are critical. Based on the three-tier management framework of virtualized network slicing, we first develop a federated learning (FL) based three-tier distributed VM anomaly detection framework, which enables distributed network slice managers to collaboratively train a global VM anomaly detection model while keeping metrics data locally. The high-dimensional, imbalanced, and distributed data features in virtualized network slicing scenarios invalidate the existing anomaly detection models. Considering the powerful ability of generative adversarial network (GAN) in capturing the distribution from complex data, we design a new multi-discriminator Bidirectional Wasserstein GAN with Gradient Penalty (BiWGAN-GP) model to learn the normal data distribution from high-dimensional resource metrics datasets that are spread on multiple VM monitors. The multi-discriminator BiWGAN-GP model can be trained over distributed data sources, which avoids high communication and computation overhead caused by the centralized collection and processing of local data. We define an anomaly score as the discriminant criterion to quantify the deviation of new metrics data from the learned normal distribution to detect abnormal behaviors arising in VMs. The efficiency and effectiveness of the proposed collaborative anomaly detection algorithm are validated through extensive experimental evaluation on a real-world dataset.
\end{abstract}

}

\maketitle

\IEEEdisplaynontitleabstractindextext

%
\IEEEpeerreviewmaketitle

\section{Introduction}
\IEEEPARstart{N}{etwork} slicing allows network operators to build multiple self-contained logical networks over a common underlying network infrastructure to accommodate the wide range of service requirements \cite{ordonez2017network,9361144}. The rise of network function virtualization (NFV) and software defined networks (SDN) facilitates the deployment of network slices through running classical network functions as software applications in virtual machines (VMs) instead of dedicated hardware \cite{yousaf2017nfv}. However, with the increasingly extensive application of VMs, their security and stability issues have drawn a wide attention \cite{8807263}. As the performance of network slices hinges on the normal running of VMs, it is crucial to detect anomalies in VMs in a timely manner to ensure the service quality of network slices.

 Anomaly detection is necessary to help find and identify abnormal behaviors in VMs before serious failures occur. The abnormal behaviors of a VM are defined as any performance degradation deviating from normal behaviors \cite{8555982}. Existing researches \cite{8231163,2018Anomaly,9110414} have shown that the abnormal behaviors of VMs usually come with a significant change in resource metrics, so it is a good way to implement anomaly detection for VMs by collecting and analyzing its multi-dimensional resource metrics data. Although there have been many interesting researches for anomaly detection, including statistical and probability methods \cite{8742597,8653385}, distance-based methods \cite{7104152,7414141}, domain-based methods \cite{miao2018distributed,9108593}, reconstruction-based methods\cite{8618315,8444085,9430511}, and information theory based methods \cite{8170290}, as classified in \cite{2014A}, detecting anomalies of VMs in virtualized network slicing environment still faces many challenges:

 Firstly, new and emerging user cases in virtualized network slicing environment have resulted in diverse types of metrics data and highly complex data structures, which are hard to accurately capture with traditional anomaly detection methods.

Secondly, the boundary between normal metrics data and abnormal ones is hard to get for three reasons: 1) Abnormal events rarely occur in real networks, so available normal and abnormal data are extremely imbalanced; 2) It is too costly to obtain enough and various kinds of abnormal metrics data in the environment with multiple demand-diverging virtual networks; 3) Novel anomalies often arise in virtualized network slicing environment.

Thirdly, substrate networks consist of multiple regions and VMs that constitute a network slice can cross multiple regions, so resource metrics data of VMs are distributed in the networks. Existing works commonly use a centralized database to aggregate all local metrics data, such as Gnocchi \cite {9482534} and Prometheus \cite {9482536,9200919}, which will compromise the communication and computation efficiency when learning a global VM anomaly detection model.

Generative adversarial network (GAN) has drawn substantial attention for its powerful ability in capturing the distribution from high-dimensional real-world data \cite{8253599}. To conquer the difficulties in capturing complex data structures, finding boundaries between normal and abnormal data and high cost of centralized training, we design a new multi-discriminator Bidirectional Wasserstein GAN with Gradient Penalty (BiWGAN-GP) to capture the distribution of normal metrics data in VMs. Compared to GAN, the multi-discriminator BiWGAN-GP includes four main improvements: 1) Replace Jensen-Shannon (JS) divergence with Wasserstein-1 distance to avoid gradient vanishing happened in GAN commonly (Wasserstein GAN (WGAN) \cite{arjovsky2017wasserstein}); 2) Besides a generator and a discriminator, an encoder is added to realize the reverse mapping  of generator to facilitate the establishment of discriminant criterion for anomaly detection (Bidirectional GAN (BiGAN) \cite{donahue2016adversarial}); 3) Gradient penalty is used to enforce the Lipschitz constraint in WGAN to solve the optimization difficulties of weight clipping (WGAN-GP \cite{Gulrajani2017}); 4) Instead of a centralized discriminator, multiple discriminators are running locally to avoid high communication and computation overhead caused by the centralized collection and processing of local data.

The management framework for virtualized network slicing is a three-tier structure, including virtualized infrastructure managers (VIMs) for VMs, network slice managers for network slices, and network controller for the whole networks. For adapting to the three-tier management framework of virtualized network slicing, we develop a federated learning (FL) based three-tier distributed anomaly detection framework to identify abnormal behaviors arising in VMs. The introduction of FL enables to collaboratively train a global VM anomaly detection model over multiple distributed datasets \cite{9082655}. The novelty of the paper lies in the combination of multi-discriminator BiWGAN-GP and FL to detect anomalies in VMs in the context of virtualized network slicing environment. Specifically, the main contributions of the paper are summarized as follows:
\begin{itemize}
\item To lower the latency when collecting data, separate databases are deployed for different regions. Besides, to ensure the isolation among different network slices, in each region, we deploy a VM monitor for a network slice to collect and store metrics data of VMs in the network slice as a local database.
\item Considering the three-tier management framework of virtualized network slicing, we develop an FL-based three-tier distributed anomaly detection framework to identify abnormal behaviors arising in VMs. The developed framework enables to collaboratively train a global VM anomaly detection model while keeping metrics data locally through the hierarchical cooperation among VM monitors, network slice managers and the network controller.
\item In view of the high-dimensional, imbalanced, and distributed data features in virtualized network slicing scenarios, we propose a new multi-discriminator BiWGAN-GP algorithm to learn the normal data distribution from high-dimensional metrics data-sets that are spread on multiple VM monitors. The proposed algorithm deploys multiple discriminators running locally in VM monitors to avoid the centralized collection and processing of local data. In addition, an anomaly score is defined as the discriminant criterion to quantify the deviation of new metrics data from the learned normal distribution to detect abnormal behaviors arising in VMs.
\item We implement extensive simulations on a real-world dataset to evaluate the efficiency and effectiveness of the proposed federated multi-discriminator BiWGAN-GP based collaborative anomaly detection algorithm. Experimental results demonstrate that the proposed algorithm can accurately detect abnormal behaviors in VMs with low communication and computation cost.
\end{itemize}

The rest of the paper is organized as follows. Section 2 presents the system model of virtualized network slicing management and the corresponding distributed anomaly detection framework. Section 3 introduces different GAN algorithms and proposes the multi-discriminator BiWGAN-GP algorithm. In Section 4, we present the training and detection processes of the three-tier FL-based distributed anomaly detection framework. The performance of the developed framework is evaluated in Section 5. Then, we provide a discussion on important related works in Section 6 and finally Section 7 concludes the paper.


\section{System Model}

\subsection{Three-tier management framework}
The network slice management and orchestration (MANO) is responsible for the control of virtualized network slicing \cite{yousaf2017nfv}. The system model of virtualized network slicing management is shown in Fig. 1, which is a three-tier structure. The specific roles of each tier are described as follows:

\begin{figure*}[htbp]
\centering
\includegraphics[width=4.5in]{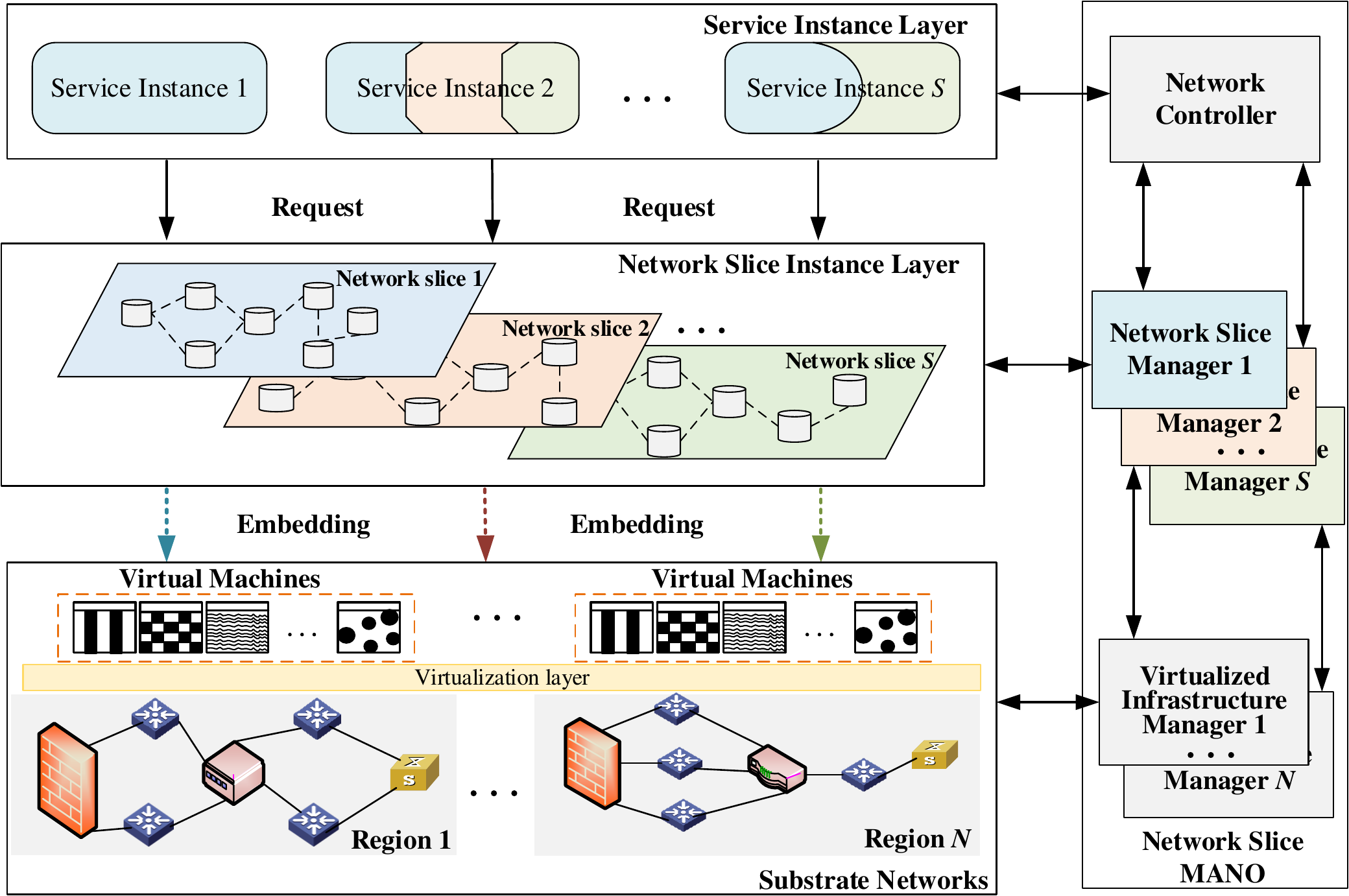}
\caption{System model of virtualized network slicing management.}
\end{figure*}

\begin{figure}[ht!]
\centering
\includegraphics[width=3.5in]{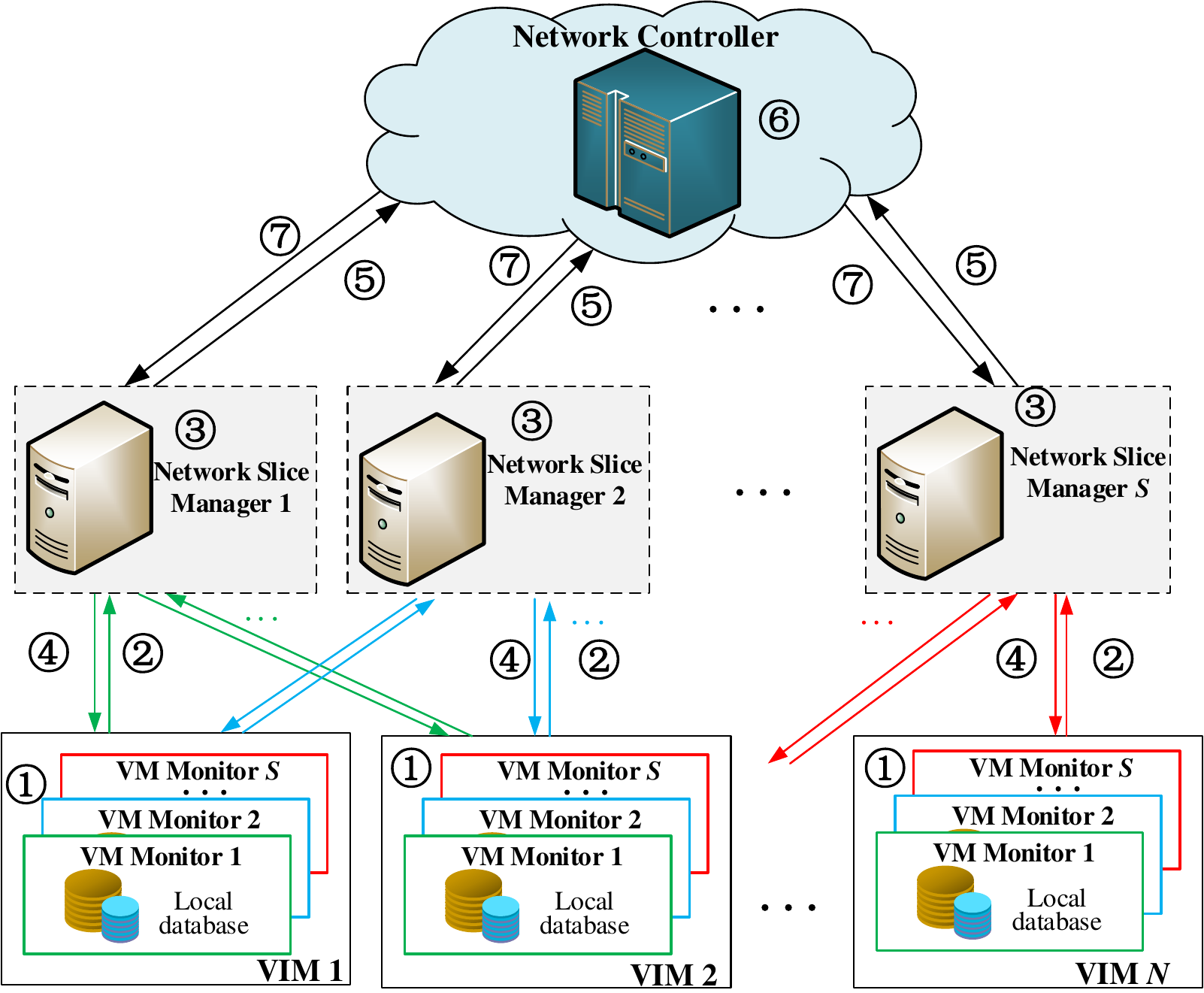}
\caption{FL-based three-tier anomaly detection framework. This framework's workflow consists of seven steps as follows: \textcircled {1} The VM monitor trains its own local discriminator model based on the local training dataset;
\textcircled {2} The VM monitor uploads error feedbacks for generator and encoder to its network slice manager;
\textcircled {3} The network slice manager updates its generator and encoder models according to the error feedbacks from all connected VM monitors, then generates fake data based on the updated generator and encoder;
\textcircled {4} The network slice manager sends the fake data to update its local discriminators on VM monitors;
\textcircled {5} After some iterations, all distributed network slice managers upload their generator and encoder models to the network controller for global aggregation;
\textcircled {6} The network controller obtains global generator and encoder models by the FedAVG algorithm \cite{konecny2017federated};
\textcircled {7} The network controller sends new global generator and encoder models to distributed network slice managers for their further updates.
The above steps are implemented repeatedly until global models converge. }

\end{figure}
\textbf{Substrate networks}: Substrate networks are divided into $N$ regions, where each region consists of various physical nodes including servers, switches and gateways. Physical nodes provide basic storage, computing and network resources required by the creation and operation of VMs. One physical node can host multiple VMs supported by the virtualization layer. VIM is responsible for the management and maintenance of each region, whose implementations include VIM connector developed in Open-Source MANO project \cite{5465112}, OpenStack and Docker.

\textbf{Network slice instances}: Network slices are defined as logical networks running on a common underlying infrastructure, mutually isolated, created on demand and with independent control and management \cite{ordonez2017network}, which is realized by distributed network slice managers. Network slice managers can be implemented by Juju charms \cite{5465112}, Tacker \cite{8456451} or Puppet \cite{yousaf2017nfv} solutions. A virtualized network slice consists of a set of virtual network functions and virtual links connecting these functions. Virtual network functions are often running on VMs as software applications \cite{8556457}.

\textbf{Service instances}: In service instance layer, service requests are received and translated into required network functions. The network controller is responsible for the orchestration of different network slices through coordinating the available resources of the whole networks, whose functionalities can be provided by OSM \cite{5465112}, OpenStack Tacker\cite{8456451} and OpenBaton \cite{6465154}.

\subsection{Three-tier distributed anomaly detection framework}
Based on the system model of virtualized network slicing management, we develop a three-tier distributed anomaly detection framework, as illustrated in Fig. 2, to detect abnormal behaviors of VMs in virtualized network slicing environment. As the abnormal behaviors of VMs usually come with significant changes in resource metrics, it is a good way to implement anomaly detection by collecting and analyzing their resource metrics data. We design a multi-discriminator BiWGAN-GP algorithm as the basic method for anomaly detection, which is elaborated in Section 3.

To lower the latency when collecting data and to ensure the isolation among network slices, in each region, we deploy a VM monitor for a network slice to collect and store metrics data of VMs in the network slice as a local database. VM monitors resort to the virtualization layer to capture resource metrics of VMs \cite{8807263, 8231163}. The resource metrics are related with the CPU usage, memory consumption, disk read/write and network input/output. To train a global VM detection model, the resource metrics data from all VMs are required. To avoid the centralized collection and processing of local data, we design an FL-based three-tier distributed anomaly detection framework to collaboratively train a global VM anomaly detection model over datasets that are spread on multiple VM monitors. The framework's workflow is illustrated in Fig. 2.

The detailed training and detection processes for the FL-based three-tier distributed anomaly detection framework are presented in Section 4.

\section{Multi-discriminator BiWGAN-GP}
The multi-discriminator BiWGAN-GP algorithm, which is a distributed form of BiWGAN-GP, trains local discriminators on local VM monitors using their own training datasets. The multi-discriminator BiWGAN-GP is designed to capture the distribution of normal metrics data in VMs.

\subsection{Existing models: GAN, BiGAN, WGAN-GP}
\emph{GAN}: GAN \cite{Goodfellow2014} consists of two models: generator $G$ and discriminator $D$. Given the real metrics data $\{ \bm{X}^r \} _{r = 1}^{N_{td} }$ as the training dataset, $G$ tries to transform a low-dimensional random noise $\bm{z}$ extracted from a known distribution $P_{\bm{z}}$ into a fake data $\bar {\bm{X}}$ (namely $\bar {\bm{X}} = G(\bm{z})$), such that  $D$ cannot distinguish $\bar {\bm{X}}$ from real data $\{ \bm{X}^r \} _{r = 1}^{N_{td} }$. Let $P_{\bm{X}}$ and  $P_{\bar {\bm{X}}}$ denote the distributions of real and fake data, and $D(\bm{X})$ denote the probability that $\bm{X}$ obeys the distribution of real data. The training objective for GAN is defined as \cite{Goodfellow2014}
\begin{equation}
\begin{gathered}
  \mathop {\min }\limits_G \;\mathop {\max }\limits_D \;V(G,D) \hfill \\
   = \mathop \mathbb{E}\limits_{{\bm{X}} \sim P_{\bm{X}} } [\log (D({\bm{X}}))] + \mathop \mathbb{E}\limits_{{\bm{z}} \sim P_{\bm{z}} } [\log (1 - D(G({\bm{z}})))] \hfill \\
   = \mathop \mathbb{E}\limits_{{\bm{X}}\sim P_{\bm{X}} } [\log (D({\bm{X}}))] + \mathop \mathbb{E}\limits_{\bar {\bm{X}} \sim P_{\bar {\bm{X}}} } [\log(1 - D(\bar {\bm{X}}))]. \hfill \\
\end{gathered}
\end{equation}

 $G$ and $D$ are updated alternately. If an optimal $D$ is obtained, the training process of $G$ theoretically leads to minimizing the JS divergence between $P_{\bm{X}}$ and $P_{\bar {\bm{X}}}$, namely,
\begin{equation}
\mathop {\min }\limits_G \;\;V(G) = 2D_{JS} (P_{\bm{X}} ||P_{{\bar {\bm{X}}}} ) - \log 4,
\end{equation}
where $D_{JS} ( \cdot || \cdot )$ represents the JS divergence. When $G$ and  $D$ are both trained into optimality, $P_{\bm{X}}=P_{\bar {\bm{X}}}$ can be achieved, which indicates that $G$ has captured the distribution of real data.

\emph{BiGAN}: Besides $G$ and $D$ in classical GAN, BiGAN also contains an encoder $E$. GAN has realized a mapping from low-dimensional random noise $\bm{z}$ in latent space to fake data $\bar {\bm{X}}$ in data space ($\bar {\bm{X}} = G(\bm{z})$), but the reverse process is unavailable. The encoder $E$ in BiGAN completes the mapping from the real data $\bm{X}$ in data space to a low-dimensional representation $\bm{f}$ in latent space ($\bm{f}=E(\bm{X})$). $D$ in BiGAN tries to distinguish between the joint pairs $(\bm{X},E(\bm{X}))$ and $(G(\bm{z}),\bm{z})$ in both data space and latent space. The training objective for BiGAN is given by
\begin{equation}
\begin{gathered}
  \mathop {\min }\limits_{G,E} \;\mathop {\max }\limits_D \;V(G,E,D) \hfill \\
   = \mathop \mathbb{E}\limits_{{\bm{X}} \sim P_{\bm{X}} } [\log (D({\bm{X}},E({\bm{X}})))] + \mathop \mathbb{E}\limits_{\bm{z} \sim P_{\bm{z}} } [\log (1 - D(G(\bm{z}),\bm{z}))] \hfill \\
   = \mathop \mathbb{E}\limits_{({\bm{X}},{\bm{f}}) \sim P_{{\bm{X}}{\bm{f}}} } [\log (D({\bm{X}},{\bm{f}})] + \mathop \mathbb{E}\limits_{({\bar {\bm{X}}},{\bm{z}}) \sim P_{{\bar {\bm{X}}}{\bm{z}}} } [\log (1 - D({\bar {\bm{X}}},z))], \hfill \\
\end{gathered}
\end{equation}
where $P_{{\bm{X}}{\bm{f}}}$ and $P_{{\bar {\bm{X}}}{\bm{z}}}$ represent the joint distributions of $({\bm{X}},{\bm{f}})$ and $({\bar {\bm{X}}},{\bm{z}})$, respectively. Given an optimal $D$, the training objective of $G$ and $E$ is also to minimize the JS divergence between $P_{{\bm{X}}{\bm{f}}}$ and $P_{{\bar {\bm{X}}}{\bm{z}}}$, namely,
\begin{equation}
\mathop {\min }\limits_{G,E} \;\;V(G,E) = 2D_{JS} (P_{{\bm{X}}{\bm{f}}} ||P_{{\bar {\bm{X}}}{\bm{z}}} ) - \log 4.
\end{equation}

The global minimum of (4) can be achieved when $P_{{\bm{X}}{\bm{f}}}=P_{{\bar {\bm{X}}}{\bm{z}}}$ is satisfied. By this time, $E$ and $G$ are each other's inverse, namely $G(E(\bm{X})) = \bm{X}$ and $E(G(\bm{z})) = \bm{z}$ \cite{donahue2016adversarial}.

\emph{WGAN-GP}:  In the early phase of GAN training, the very little overlap between  $P_{\bm{X}}$ and $P_{\bar {\bm{X}}}$  will cause their JS divergence being a constant and the gradient vanishing, which will terminate the training exceptionally. To solve it, WGAN \cite{arjovsky2017wasserstein} replaces the JS divergence with Wasserstein-1 distance, which is computed by
\begin{equation}
W_1 (P_{\bm{X}} ||P_{\bar {\bm{X}}} ) = \mathop {\sup }\limits_{f \in \mathbb{L}_1 } \{ \mathop \mathbb{E}\limits_{{\bm{X}} \sim P_{\bm{X}} } [f({\bm{X}})] - \mathop \mathbb{E}\limits_{\bar {\bm{X}} \sim P_{\bar {\bm{X}}} } [f(\bar {\bm{X}})]\},
\end{equation}
where $\mathbb{L}_1$ is  the set of 1-Lipschitz functions, which is satisfied through weight clipping \cite{arjovsky2017wasserstein}. Due to the optimization difficulties of weight clipping, \cite{Gulrajani2017} proposed WGAN with gradient penalty (WGAN-GP) to enforce the Lipschitz constraint. WGAN-GP uses $D$ with the norm of its gradient penalty to approximate the function $f \in \mathbb{L}_1$  in WGAN. Therefore, $D$ is trained based on the following objective:
\begin{equation}
\begin{gathered}
  \mathop {\max }\limits_D \;V(D) = \mathop \mathbb{E}\limits_{{\bm{X}} \sim P_{\bm{X}} } [D(X)] - \mathop \mathbb{E}\limits_{{\bar {\bm{X}}} \sim P_{\bar {\bm{X}}} } [D({\bar {\bm{X}}})] \hfill \\
  \;\;\;\;\;\;\;\;\;\;\;\;\;\;\;\;\; + \eta \mathop \mathbb{E}\limits_{\widehat{{\bm{X}}} \sim P_{\widehat {\bm{X}}} } [(||\nabla _{\widehat{{\bm{X}}}} D(\widehat{{\bm{X}}})||_2  - 1)^2 ], \hfill \\
\end{gathered}
\end{equation}
where $\eta$ is the penalty coefficient and $\widehat{{\bm{X}}}$ is defined to uniformly sample along straight lines between pairs of points extracted from the distributions $P_{\bm{X}}$ and $P_{\bar {\bm{X}}}$. Therefore, $\widehat{{\bm{X}}}$ is the weighted sum of $\bm{X}$ and $\bar {\bm{X}}$: $\widehat{{\bm{X}}} = \varepsilon \bm{X} + (1-\varepsilon)\bar {\bm{X}}$, where $\varepsilon \in U[0,1]$. $G$ is trained through
\begin{equation}
\mathop {\min }\limits_G \;\;V(G) =  - \mathop \mathbb{E}\limits_{\bar {\bm{X}} \sim P_{\bar {\bm{X}}} } [D(\bar {\bm{X}})] =  - \mathop \mathbb{E}\limits_{\bm{z}\sim P_{\bm{z}} } [D(G(\bm{z}))].
\end{equation}

\begin{figure}[t]
\centering
\includegraphics[width=3.5in]{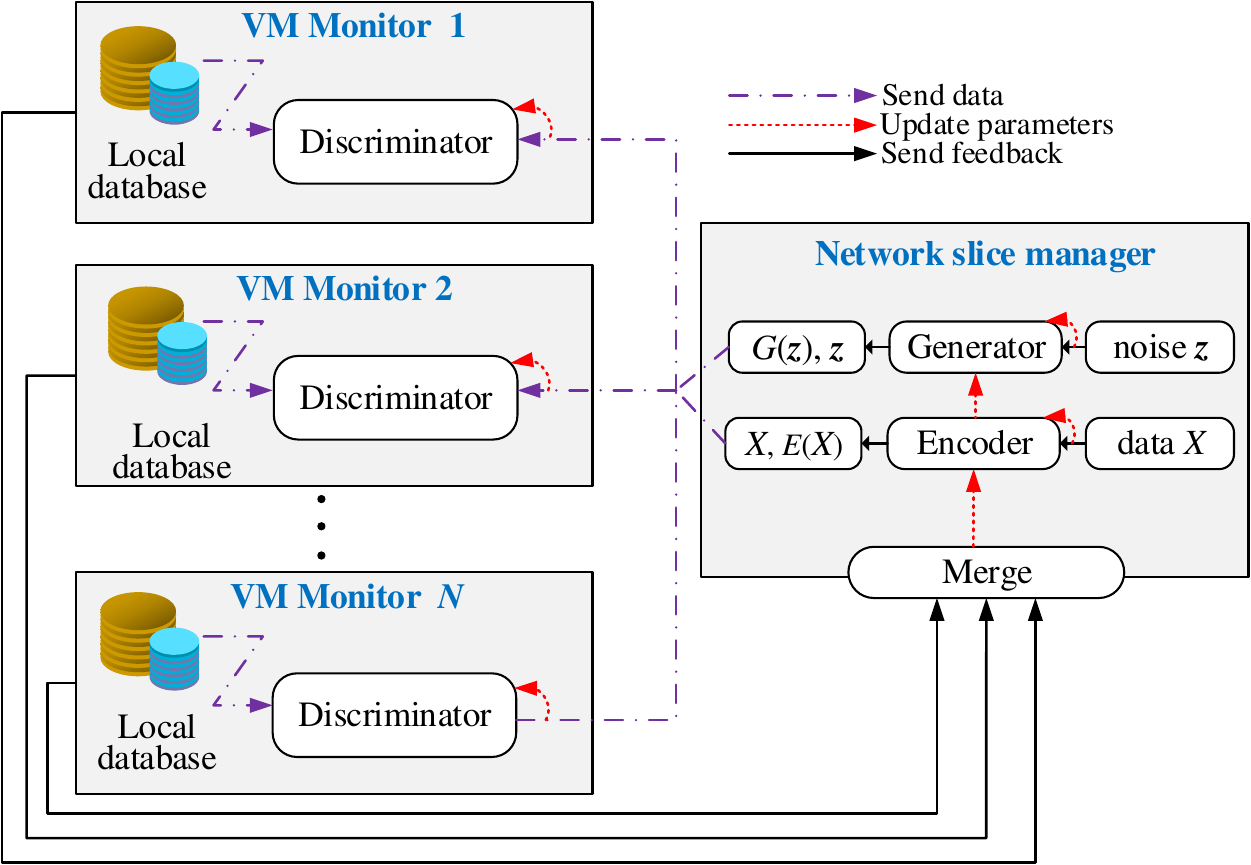}
\caption{Multi-discriminator BiWGAN-GP framework. }
\end{figure}

\subsection{BiWGAN-GP and its distributed form}
 We propose BiWGAN-GP to combine the strengths of BiGAN and WGAN-GP.  After obtaining $(\bm{X},\bm{f})$ and $({\bar {\bm{X}}}, \bm{z})$, $(\widehat{{\bm{X}}},\widehat{{\bm{z}}})$ is calculated through  $(\widehat{{\bm{X}}},\widehat{{\bm{z}}}) = \varepsilon(\bm{X},\bm{f})+ (1 - \varepsilon)({\bar {\bm{X}}}, \bm{z})$ similar to the calculation of $\widehat{{\bm{X}}}$. In BiWGAN-GP, $D$ is trained through
\begin{equation}
\begin{gathered}
  \mathop {\max }\limits_D \;V(D) = \mathop \mathbb{E}\limits_{({\bm{X}},{\bm{f}}) \sim P_{{\bm{Xf}}} } [D({\bm{X}},{\bm{f}})] - \;\;\mathop \mathbb{E}\limits_{({\bar {\bm{X}}},{\bm{z}}) \sim P_{{\bar {\bm{X}}}{\bm{z}}} } [D({\bar {\bm{X}}},{\bm{z}})]  \hfill \\
  \;\;\;\;\;\;\;\;\;\;\;\;\;\;\;\;\;\;+\eta \mathop \mathbb{E}\limits_{(\widehat {\bm{X}},\widehat {\bm{z}}) \sim P_{\widehat {\bm{X}} \widehat {\bm{z}}} } [(||\nabla _{(\widehat {\bm{X}},\widehat {\bm{z}})} D((\widehat {\bm{X}},\widehat {\bm{z}}))||_2  - 1)^2 ], \hfill \\
\end{gathered}
\end{equation}
where the definition of $P_{\widehat {\bm{X}} \widehat {\bm{z}}}$  is similar to $P_{\widehat {\bm{X}}}$. After training $D$, the Wasserstein-1 distance $W_1 (P_{{\bm{Xf}}} || P_{{\bar {\bm{X}}}{\bm{z}}} )$ between $P_{{\bm{Xf}}}$ and $P_{{\bar {\bm{X}}}{\bm{z}}}$  can be expressed as
\begin{equation}
\begin{gathered}
  W_1 (P_{{\bm{Xf}}} || P_{{\bar {\bm{X}}}{\bm{z}}} ) =  \hfill \\
  \;\;\;\;\;\;\;\mathop \mathbb{E}\limits_{({\bm{X}},{\bm{f}}) \sim P_{{\bm{Xf}}} } [D({\bm{X}},{\bm{f}})] - \mathop \mathbb{E}\limits_{({\bar {\bm{X}}},{\bm{z}}) \sim P_{{\bar {\bm{X}}}{\bm{z}}} } [D({\bar {\bm{X}}},{\bm{z}})]. \hfill \\
\end{gathered}
\end{equation}

The training objective of $G$ and $E$  is to minimize $W_1 (P_{{\bm{Xf}}} || P_{{\bar {\bm{X}}}{\bm{z}}} )$, namely,
\begin{equation}
\mathop {\min }\limits_{G,E} \;\;V(G,E) = W_1 (P_{{\bm{Xf}}} || P_{{\bar {\bm{X}}}{\bm{z}}} ).
\end{equation}

A multi-discriminator BiWGAN-GP is designed to learn the distribution of all metrics data within a network slice. In multi-discriminator BiWGAN-GP, a pair of generator and encoder is trained on a network slice manager and $N$ discriminators are trained on VM monitors.

The multi-discriminator BiWGAN-GP framework is illustrated in Fig. 3. The network slice manager selects and sends the generated data to each VM monitor, and the discriminators in VM monitors try to distinguish generated data from local real data. Therefore, the optimization objective of multi-discriminator BiWGAN-GP is defined as
\begin{equation}
\begin{gathered}
  \mathop {\min }\limits_{G,E} \;\mathop {\max }\limits_D \;V(G,E,D) =  \hfill \\
  \frac{1}
{N}\sum\limits_{n = 1}^N {\{ \mathop \mathbb{E}\limits_{({\bm{X}},{\bm{f}}) \sim P_{{\bm{Xf}}} } [D_n ({\bm{X}},{\bm{f}})] - \mathop \mathbb{E}\limits_{({\bar {\bm{X}}},{\bm{z}}) \sim P_{{\bar {\bm{X}}}{\bm{z}}} } [D_n ({\bar {\bm{X}}},{\bm{z}})]}  \hfill \\
   + \mathop \mathbb{E}\limits_{(\widehat {\bm{X}},\widehat {\bm{z}}) \sim P_{\widehat {\bm{X}} \widehat {\bm{z}}} } [(||\nabla _{(\widehat {\bm{X}},\widehat {\bm{z}})} D_n ((\widehat {\bm{X}},\widehat {\bm{z}}))||_2  - 1)^2 ]\}.  \hfill \\
\end{gathered}
\end{equation}

\subsection{Modeling the multi-discriminator BiWGAN-GP}
As the available training dataset of metrics is time series, we choose Vanilla Long Short-Term Memory Network (VLSTM) as the basic model of generator  $G$ and encoder $E$, which has shown superior performance on time-series-related tasks as a kind of recurrent neural networks \cite{7508408}. Given $[{\bm y}_1 ,...,{\bm y}_t ]$ as input, then for each time epoch $\tau$ ($\tau  = 1,...,t$), we have
\begin{equation}
\begin{gathered}
  {\bm i}_\tau   = \sigma ({\bm w}_i .[\bm {y}_\tau  ,\;\bm {h}_{\tau  - 1} ,\;\bm{c}_{\tau  - 1} ] + \bm{b}_i ), \hfill \\
  \bm{f}_\tau   = \sigma (\bm{w}_f .[y_\tau  ,\;\bm{h}_{\tau  - 1} ,\;\bm{c}_{\tau  - 1} ] + \bm{b}_f ), \hfill \\
  \bm{c}_\tau   = \bm{f}_\tau   \odot \bm{c}_{\tau  - 1}  + \bm{i}_\tau   \odot \tanh (\bm{w}_z .[\bm {y}_\tau  ,\;\bm{h}_{\tau  - 1} ] + \bm{b}_z ), \hfill \\
  \bm{o}_\tau   = \sigma (\bm{w}_o .[\bm {y}_\tau  ,\;\bm{h}_{\tau  - 1} ,\;\bm{c}_\tau  ] + \bm{b}_o ), \hfill \\
  \bm{h}_\tau   = \bm{o}_\tau   \odot \tanh (\bm{c}_\tau  ), \hfill \\
\end{gathered}
\end{equation}
where ${\bm i}_\tau$, $\bm{f}_\tau $ and $\bm{o}_\tau$ are the input gate, forget gate and output gate, respectively. $\bm{c}_\tau$ denotes the memory unit and $\bm{h}_\tau$ denotes the hidden state. $\bm{w}_*$ and $\bm{b}_*$ represent weights and bias, respectively. The symbol $\sigma ( \cdot )$ represents the sigmoid function and $\odot$ denotes the operation of element-wise multiplication.

\emph{Build} $G$: We build the generator $G$ into a three-layer structure, including two VLSTM layers and one fully connected layer activated by linear. In network slice managers, after extracting $\bm{z}$ from $P_{\bm{z}}$, $G$ maps $\bm{z}$ into the fake data $\bar {\bm X}$
\begin{equation}
\bar {\bm X} = G(\bm{z};\;[\bm {w}_G, \bm{b}_G ]),
\end{equation}
where $\bm {w}_G$ and $\bm{b}_G$ denote weights and biases of $G$, which compose model parameters $\theta _G$.

\emph{Build} $E$: As encoder $E$ is the inverse function of $G$, it is also a three-layer structure with two VLSTM layers and one fully connected layer activated by linear, but they are grouped in a fully reverse order to  $G$. $E$ maps the real data $\bm{X}$ to a low-dimensional representation $\bm{f}$
\begin{equation}
\bm {f} = E(\bm{X};\;[\bm {w}_E, \bm{b}_E ]),
\end{equation}
where $\bm {w}_E$ and $\bm{b}_E$ denote weights and biases of $E$, which compose model parameters $\theta _E$.

\emph{Build} $D_n$: We build the discriminator $D_n\;(n \in N)$ with three fully connected layer, where the last layer is sigmoid activated to get a probability in range $(0,1)$. The discriminator $D_n$ in each VM monitor maps $({\bm X}_n,{\bm f}_n)$ and $({\bar {{\bm X}}_n},{\bm z}_n)$ to a scalar $d$
\begin{equation}
d = D_n \left\{ {\left[ {({\bm X}_n,{\bm f}_n),\;({\bar {{\bm X}}_n},{\bm z}_n)} \right];[\bm {w}_{D_n } ,\bm {b}_{D_n } }] \right\},
\end{equation}
where $\bm {w}_{D_n }$ and $\bm {b}_{D_n }$ denote weights and biases of $D_n$, which compose model parameters $\theta _{D_n}$.

The training of multi-discriminator BiWGAN-GP algorithm for a network slice is iteratively performed by VM monitors and the network slice manager.  Next, we separately specify the training of $D_n$ ($n \in N$) on each VM monitor and the training of $G$ and $E$ on the network slice manager.

\emph{Training on VM monitors}: Each VM monitor hosts a discriminator $D_n$ with its parameters $\theta_{D_n}$. In each iteration, a VM monitor contains two functions: 1) Receive the generated data to update its discriminator; 2) Calculate the error feedbacks for generator and encoder updates.

1) \emph{Discriminator Update}: Each VM monitor receives the generated data pairs $({\bar {{\bm X}}_n},{\bm z}_n)$ and the low-dimensional representation $\bm{f}_n$ of $\bm{X}_n$  from the network slice manager, whose batch sizes are all $M$. Then, the VM monitor performs $K$ training iterations on $D_n$. In the $k$th local iteration, for each batch $({\bm X}_n^m,{\bm f}_n^m)$ and $({\bar {{\bm X}}_n^m},{\bm z}_n^m)$  in $({\bm X}_n,{\bm f}_n)$ and $({\bar {{\bm X}}_n},{\bm z}_n)$ , the loss function of $D_n$ is calculated by
\begin{equation}
\begin{gathered}
  L_{D_n }^m  =  - \{ D_n({\bm X}_n^m ,{\bm f}_n^m ) - D_n({\bar {\bm X}}_n^m ,{\bm z}_n^m )+ \hfill \\
    \eta [||\Delta _{(\widehat {\bm X}_n^m ,\widehat {\bm z}_n^m )} D_n(\widehat {\bm X}_n^m ,\widehat {\bm z}_n^m )||_2  - 1]^2 \}. \hfill \\
\end{gathered}
\end{equation}

Accordingly, the gradients $\Delta \theta _{D_n }$ are calculated by
\begin{equation}
\Delta \theta _{D_n }  = \frac{1}
{M}\sum\limits_{m = 1}^M {\frac{{\partial L_{D_n }^m }}
{{\partial \theta _{D_n } }}}.
\end{equation}

We update the parameters $\theta _{D_n }$ of discriminator $D_n$  using the Adam optimizer method
\begin{equation}
\theta _{D_n }  \leftarrow {\text{Adam}}(\Delta \theta _{D_n } ,\theta _{D_n } ,\alpha ,\beta _1 ,\beta _2 ),
\end{equation}
where $\alpha$, $\beta _1$ and $\beta _2$ are hyperparameters of Adam optimizer.

2) \emph{Error Feedbacks Calculation}: To update the parameters of  $G$ and $E$ in the network slice manager, each VM monitor calculates the error feedbacks $F_G^n$ and $F_E^n$ based on the local loss function $L_{EG}^n$ and $D_n$ after $K$ training iterations. The local loss function $L_{EG}^n$  for $G$ and $E$ in each VM monitor is computed as
\begin{equation}
L_{EG}^n  = \frac{1}
{M}\sum\limits_{m = 1}^M {\{ D_n({\bm X}_n^m ,{\bm f}_n^m ) - D_n({\bar {\bm X}}_n^m ,{\bm z}_n^m)\} }.
\end{equation}

The error feedback $F_E^n$ of $E$ consists of $M$ vectors $\{ \bm {e}_n^1 ,...,\bm{e}_n^M \}$, where $\bm{e}_n^m$ is defined as
\begin{equation}
e_n^m  = \frac{{\partial L_{EG}^n }}
{{\partial ({\bm X}_n^m ,{\bm f}_n^m )}}.
\end{equation}

Similarly, the error feedback $F_G^n$ of $G$ also consists of $M$ vectors $\{ \bm {g}_n^1 ,...,\bm{g}_n^M \}$, where $\bm{g}_n^m$ is defined as
\begin{equation}
g_n^m  = \frac{{\partial L_{EG}^n }}
{{\partial ({\bar {\bm X}}_n^m ,{\bm z}_n^m )}}.
\end{equation}

$F_E^n$  and $F_G^n$ obtained in each VM monitor will be sent to the network slice manager for $G$ and $E$ updates.

\emph{Training on the network slice manager}: The network slice manager hosts $G$ and $E$, which contains two functions in each iteration: 1) Generate and encode data using the generator and encoder; 2) Receive error feedbacks to update the generator and encoder.

1) \emph{Data Generation}: The network slice manager first receives the set of training batches $\{ \bm {X}_n \} _{n = 1}^N$ from $N$ VM monitors. $E$ turns  $\{ \bm {X}_n \} _{n = 1}^N$ into a set of low-dimensional representations $\{ \bm {f}_n \} _{n = 1}^N$. In addition, $G$ transforms a set of low-dimensional random noise $\{ \bm {z}_n \} _{n = 1}^N$ extracted from a known distribution $P_{\bm z}$ into a set of fake data $\{ \bar {\bm X}_n \} _{n = 1}^N$. Then, the network slice manager sends $\bm{f}_n$, $\bm{z}_n$ and $\bar {\bm X}_n$ to each VM monitor to train $D_n$ ($n \in N$).

2) \emph{$G$ and $E$ Updates}: After receiving error feedbacks  $F_E^n$  and $F_G^n$ from each VM monitor, the network slice manager updates parameters $\theta_G$ and $\theta_E$ of  $G$ and $E$. We use the averaging operation to aggregate error feedbacks from all VM monitors as it is the most common method to merge feedbacks updated in parallel \cite{8821025}. Specifically, the gradients $\Delta \theta _G$ of $\theta _G$ are deduced from feedbacks $F_G^n$ ($n \in N$) by
\begin{equation}
\begin{gathered}
  \Delta \theta _G^l  = \frac{1}
{{MN}}\sum\limits_{n = 1}^N {\sum\limits_{m = 1}^M {\frac{{\partial L_{EG}^n }}
{{\partial \theta _G^l }}} }  \hfill \\
  \;\;\;\;\;\;\; = \frac{1}
{{MN}}\sum\limits_{n = 1}^N {\sum\limits_{m = 1}^M {\frac{{\partial L_{EG}^n }}
{{\partial (\bar {\bm X}_n^m,{\bm z}_n^m )}}} \frac{{\partial (\bar {\bm X}_n^m,{\bm z}_n^m  )}}
{{\partial \theta _G^l }}}  \hfill \\
  \;\;\;\;\;\;\; = \frac{1}
{{MN}}\sum\limits_{n = 1}^N {\sum\limits_{m = 1}^M {\bm {g}_n^m } \frac{{\partial (\bar {\bm X}_n^m,{\bm z}_n^m  )}}
{{\partial \theta _G^l }}},  \hfill \\
\end{gathered}
\end{equation}
where $\Delta \theta _G^l$ is the $l$-th ($l \in L$) element of gradients $\Delta \theta _G$. After the gradients are calculated, the network slice manager updates parameters $\theta _G$ using Adam optimizer
\begin{equation}
\theta _G  \leftarrow \text{Adam}\{(\Delta \theta _G^1 ,...,\Delta \theta _G^L) ,\theta _G ,\alpha ,\beta _1. ,\beta _2 \}.
\end{equation}

Similarly, the gradients $\Delta \theta _E$ of parameters $\theta _E$ are deduced from all feedbacks $F_E^n$ ($n \in N$) by
\begin{equation}
\begin{gathered}
  \Delta \theta _E^l  = \frac{1}
{{MN}}\sum\limits_{n = 1}^N {\sum\limits_{m = 1}^M {\frac{{\partial L_{EG}^n }}
{{\partial \theta _E^l }}} }  \hfill \\
  \;\;\;\;\;\;\; = \frac{1}
{{MN}}\sum\limits_{n = 1}^N {\sum\limits_{m = 1}^M {\frac{{\partial L_{EG}^n }}
{{\partial ({\bm X}_n^m ,{\bm f}_n^m )}}} \frac{{\partial ({\bm X}_n^m ,{\bm f}_n^m )}}
{{\partial \theta _E^l }}}  \hfill \\
  \;\;\;\;\;\;\; = \frac{1}
{{MN}}\sum\limits_{n = 1}^N {\sum\limits_{m = 1}^M {\bm {e}_n^m } \frac{{\partial ({\bm X}_n^m ,{\bm f}_n^m )}}
{{\partial \theta _E^l }}},  \hfill \\
\end{gathered}
\end{equation}
where $\Delta \theta _E^l$ is the $l$-th ($l \in L$) element of gradients $\Delta \theta _E$. As $G$ and $E$  are each other's inverse, $\Delta \theta _E$ has the same number of elements with $\Delta \theta _G$. Then, the network slice manager updates parameters $\theta _E$ using Adam optimizer
\begin{equation}
\theta _E  \leftarrow {\text{Adam}}\{(\Delta \theta _E^1 ,...,\Delta \theta _E^L ) ,\theta _E ,\alpha ,\beta _1 ,\beta _2 \}.
\end{equation}

The detailed steps of multi-discriminator BiWGAN-GP algorithm are presented in Algorithm 1.

\begin{algorithm}[t]
\caption{Multi-discriminator BiWGAN-GP algorithm}
\begin{algorithmic}[1]
\REQUIRE The number of iterations $K$ between two $E$ and $G$ iterations. Batch size $M$. Adam hyperparameters $\alpha ,\;\beta _1 ,\;\beta _2$. The prior distribution $P_{\bm z}$.  The number of training iterations $I$.
\renewcommand{\algorithmicrequire}{ \textbf{Input:}}
\REQUIRE Initialized model parameters: $\theta _G$, $\theta _E$, $\theta _{D_n}$ ($n \in N$), and distributed training datasets $\cup _{n = 1}^N \{ \bm {X}_n^r \} _{r = 1}^{N_{td} }$.
\ENSURE Converged model parameters: $\theta _G$, $\theta _E$, $\theta _{D_n}$($n \in N$).
\FOR {$i = 1:I$}
\STATE $\rhd$ \textbf{Update $D_n$ ($n \in N$) on VM monitors}
\STATE Sample $\bm{X}_n$ from its own training dataset $\{ \bm{X}_n^r \} _{r = 1}^{N_{td} }$ and send it to the network slice manager
\STATE Receive $({\bar {\bm X}}_n,{\bm z}_n)$ and $\bm {f}_n$ from the network slice manager to obtain the data pairs $({\bm X}_n,{\bm f}_n)$ and $({\bar {\bm X}}_n,{\bm z}_n)$     
\FOR {$k = 1:K$}
\FOR {$m = 1:M$}
\STATE Calculate $(\widehat{{\bm{X}}}_n^m,\widehat{{\bm{z}}}_n^m) = \varepsilon(\bm{X}_n^m,\bm{f}_n^m)+ (1 - \varepsilon)({\bar {\bm{X}}}_n^m, \bm{z}_n^m)$
\STATE Calculate $L_{D_n }^m$ according to (16)
\IF {$k==K$}
\STATE Calculate $L_{EG}^n$ according to (19)
\STATE Calculate elements $\bm{e}_n^m$ and $\bm{g}_n^m$ of error feedbacks according to (20) and (21)
\ENDIF
\ENDFOR
\STATE Update gradients $\Delta \theta _{D_n}$  and $D_n$'s parameters $\theta _{D_n}$ according to (17) and (18)
\ENDFOR
\STATE Send the error feedbacks $F_G^n  = \{ \bm{g}_n^1 ,...,\bm{g}_n^M \}$ and $F_E^n  = \{ \bm{e}_n^1 ,...,\bm{e}_n^M \}$ to the network slice manager
\STATE $\rhd$ \textbf{Update $E$ and $G$ on the network slice manager}
\STATE  Receive a set of training batches $\{ \bm {X}_n \} _{n = 1}^N$ from $N$ VM monitors
\STATE  Sample a set of noise $\{ \bm{z}_n \} _{n = 1}^N  \sim P_{\bm z}$
\STATE  Obtain $\{ \bm{f}_n \} _{n = 1}^N  \leftarrow E(\{ \bm{X}_n \} _{n = 1}^N )$ and $\{ \bar {\bm X}_n \} _{n = 1}^N  \leftarrow G(\{ \bm{z}_n \} _{n = 1}^N )$
\STATE  Send $\bm{f}_n$, $\bm{z}_n$ and $\bar {\bm{X}}_n$ to each VM monitor
\STATE Deduce gradients $\Delta \theta _G$ of $G$  from all feedbacks $F_G^n$ ($n \in N$) based on (22), then update $G$'s parameters $\theta_G$ according to (23)
\STATE Deduce gradients $\Delta \theta _E$ of $E$ from all feedbacks $F_E^n$ ($n \in N$) based on (24), then update $E$'s parameters $\theta_E$ according to (25)
\ENDFOR
\end{algorithmic}
\end{algorithm}

\begin{figure*}[htbp]
\centering
\includegraphics[width=5.3in]{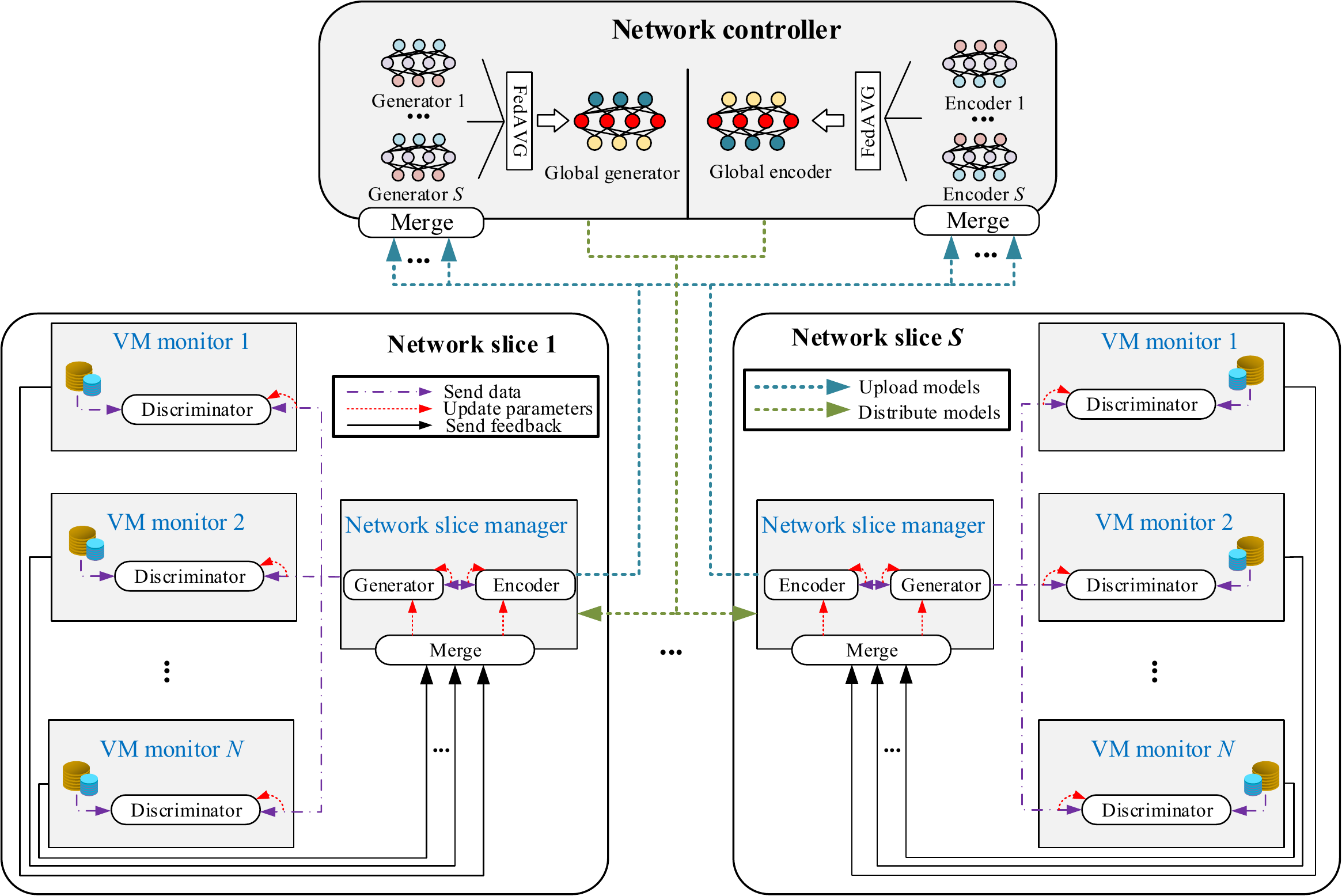}
\caption{FL-based three-tier distributed anomaly detection framework.}
\end{figure*}

\section{FL-based three-tier distributed anomaly detection framework}
The proposed multi-discriminator BiWGAN-GP algorithm is adapted to the management of VMs within a network slice. To extend it to the entire network, FL is introduced to the multi-discriminator BiWGAN-GP algorithm to develop an FL-based three-tier distributed anomaly detection framework as shown in Fig. 4.  From the perspective of a network slice, local training data in each VM monitor are utilized to train the multi-discriminator BiWGAN-GP model. After some iterations, network slice managers will send the local models of generators and encoders to the network controller for global aggregation, such that we can achieve a global VM anomaly detection model through the hierarchical cooperation among VM monitors, network slice managers and the network controller.

\subsection{Training}
The training of FL-based multi-discriminator BiWGAN-GP algorithm on network slices and the network controller are separately specified as follows:

\emph{Network slices}: A multi-discriminator BiWGAN-GP model is implemented within a network slice, which contains one pair of generator and encoder and $N$ discriminators. $N$ Discriminators are trained on VM monitors and local generators  $G_s$ and encoders $E_s \;(s \in S)$ are trained on network slice managers, where $S$ represents both the number and set of network slices. Network slice managers first receive initial global models $G^{global}$ and $E^{global}$  of generator and encoder from the network controller as the initial local ones. In a network slice $s$, the network slice manager and VM monitors exchange data and parameters every $K$ iterations for the training of  $D_n$ ($n \in N$), $G_s$  and $E_s$. After $L$ iterations, each network slice manager sends model parameters $\theta _{G_s }$ and $\theta _{E_s }$ to the network controller for global aggregation. After the averaging operation in network controller, it sends new global models $G^{global}$ and $E^{global}$ to network slice managers for updating local $G_s$ and $E_s\;(s \in S)$  using the Adam optimizer method
\begin{equation}
\begin{gathered}
  \theta _{G_s }  \leftarrow {\text{Adam}}(\Delta \theta _{G_s } ,\theta _G^{global} ,\alpha ,\beta _1 ,\beta _2 ), \hfill \\
  \theta _{E_s }  \leftarrow {\text{Adam}}(\Delta \theta _{E_s } ,\theta _E^{global} ,\alpha ,\beta _1 ,\beta _2 ), \hfill \\
\end{gathered}
\end{equation}
where $\theta _G^{global}$ and $\theta _E^{global}$ are global model parameters of $G^{global}$ and $E^{global}$.

\emph{Network controller}: Network controller is responsible for the global aggregation of $G_s$ and $E_s \;(s \in S)$. Network controller contains two main functions: (1) Initialize global models and send them to network slice managers; (2) Aggregate all model parameters $\theta _{G_s }$ and $\theta _{E_s } \;(s \in S)$ uploaded by network slice managers, then average the received parameters and send global models $G^{global}$ and $E^{global}$ back to network slice managers. $G^{global}$ and $E^{global}$ are updated using averaging operation (i.e., FedAVG algorithm \cite{konecny2017federated}), which is given by
\begin{equation}
\theta _G^{global}  = \frac{{\sum\nolimits_{s = 1}^S {Q_s \theta _{G_s } } }}
{Q},\;\theta _E^{global}  = \frac{{\sum\nolimits_{s = 1}^S {Q_s \theta _{E_s } } }}
{Q},
\end{equation}
where $Q_s$ is the number of training data in network slice $s$  ($s \in S$), and $Q = \sum\nolimits_{s = 1}^S {Q_s }$ is the total number of training data in all network slices.

The above steps are implemented repeatedly until the global models $G^{global}$ and $E^{global}$ reach optimal convergence. In addition, the network environment is not static but time-varying. The trained model may become inaccurate with time. Therefore, the proposed FL-based multi-discriminator BiWGAN-GP algorithm still needs to be retrained periodically on the newly collected data.

The proposed algorithm can be extended to the dynamic network environment. If there are new network slices joining the FL, we can transfer the well-trained model as their initial models to detect possible abnormal behaviors in current period, and then update their anomaly detection model in the next retraining period.

\subsection{Detection}
After the generator and encoder are well trained, the copy of them will be sent to each VM monitor for VM anomaly detection. To measure the abnormality of VM $v$ $(v \in V)$, we introduce an anomaly score  $A({\bm X}_v)$ for the new metrics data ${\bm X}_v$. The anomaly score $A({\bm X}_v)$ is defined as a weighted combination of the reconstruction loss $L_{EG}$ and the discriminator loss $L_D$, which is calculated by
\begin{equation}
A({\bm X}_v) = \gamma L_{EG} ({\bm X}_v) + (1 - \gamma )L_D ({\bm X}_v),
\end{equation}
where $\gamma$ is the weighted coefficient, $L_{EG} ({\bm X}_v) = ||{\bm X}_v - G(E({\bm X}_v))||_1$ and $L_D ({\bm X}_v) = \sigma (D({\bm X}_v),E({\bm X}_v),1)$. In $L_D ({\bm X}_v)$, $\sigma$ is the sigmoid cross entropy, which defines the discriminator's confidence that ${\bm X}_v$ obeys the real data distribution \cite{Schlegl2017}. For new metrics data ${\bm X}_v$  of VM $v$, its VM monitor calculates the anomaly score $A({\bm X}_v)$ based on its own discriminator and the received generator and encoder. If $A({\bm X}_v)$ is larger than a preset threshold, VM $v$ will be determined as abnormal.

The detailed steps of FL-based multi-discriminator BiWGAN-GP algorithm are presented in Algorithm 2.

\begin{algorithm}[t]
\caption{FL-based multi-discriminator BiWGAN-GP}
\begin{algorithmic}[1]
\renewcommand{\algorithmicrequire}{ \textbf{Hyperparameters:}}
\REQUIRE The number of iterations $K$ between two $E$ and $G$ iterations. Batch size $M$. Adam hyperparameters $\alpha ,\;\beta _1 ,\;\beta _2$. The number of training iterations $I$. The number of local iterations $L$ before aggregation. The value of $threshold$ for anomaly score.
\renewcommand{\algorithmicrequire}{ \textbf{Input:}}
\REQUIRE Initialized model parameters: $\theta _G^{global}$, $\theta _E^{global}$, $\theta _{D_n}$ ($n \in N$), and distributed training datasets $\cup _{n = 1}^N \{ \bm {X}_n^r \} _{r = 1}^{N_{td} }$ for all $s \in S$, new  metrics data ${\bm X}_v$ of VM $v$ ($v \in V$).
\ENSURE Converged model parameters: $\theta _G^{global}$, $\theta _E^{global}$  and $\theta _{D_n}$ ($n \in N$) for all $s \in S$, the abnormality of VM $v$ .
\FOR[$\rhd$ \textbf{Training process}]{$i = 1:I$}
\STATE $\rhd$ \textbf{Update $D_n$ ($n \in N$) for all $s \in S$ on VM monitors}
\STATE Implement steps (3)-(16) in \textbf{Algorithm 1}
\STATE $\rhd$ \textbf{Update $E_s$ and $G_s$ ($s \in S$) on network slice managers}
\STATE Receive initial global parameters $\theta _G^{global}$ and $\theta _E^{global}$ from the network controller as initial local parameters $\theta _{G_s }$ and $\theta _{E_s }$
\STATE  Implement steps (18)-(24) in \textbf{Algorithm 1}
\IF {$i\;\bmod \;L =  = 0$}
\STATE Send local parameters $\theta _{G_s }$ and $\theta _{E_s }$ to the network controller for aggregation
\STATE Receive updated global parameters $\theta _G^{global}$ and $\theta _E^{global}$ and use them update local  $G_s$ and $E_s$ models in each network slice $s$ according to (26)
\ENDIF
\STATE $\rhd$ \textbf{Update global models $G^{global}$ and $E^{global}$ on the network controller}
\STATE Initialize global model parameters  $\theta _G^{global}$ and $\theta _E^{global}$ and sends them to all network slice managers
\STATE Update global model parameters $\theta _G^{global}$ and $\theta _E^{global}$ by averaging $\theta _{G_s }$ and $\theta _{E_s }$ ($s \in S$) sent from network slice managers every $L$ iterations according to (27)
\ENDFOR
\STATE For new metrics data ${\bm X}_v$ of VM $v$, calculate its anomaly score $A({\bm X}_v)$ based on its corresponding $G$, $E$ and $D$ according to (28)\;\;\;\;\; \COMMENT{$\rhd$ \textbf{Detection process}}
\IF {$A({\bm X}_v)> threshold$}
\STATE VM $v$ is determined as abnormal
\ELSE
\STATE VM $v$ is determined as normal
\ENDIF
\end{algorithmic}
\end{algorithm}

\section{Performance Evaluation}
In this section, we evaluate the performance of the proposed FL-based multi-discriminator BiWGAN-GP algorithm in terms of efficiency and effectiveness through extensive experimental evaluation, and compare it with GAN, BiGAN, WGAN-GP, centralized BiWGAN-GP, standalone BiWGAN-GP, and multi-discriminator BiWGAN-GP algorithms.

\subsection{Experiment setup}
We implement the centralized, standalone, distributed and the developed FL-based distributed anomaly detection framework in Python language using Pytorch package on Intel(R) Core(TM) i7-6700H CPU with 16GB RAM. In our implementations, we emulate 1 network controller, 3 network slice managers, and 9 VM monitors, and use socket to realize the information exchange among them.

1) \emph{Datasets}: We choose the VNFDataset (virtual IP Multimedia IP system) \footnote{VNFDataset: virtual IP Multimedia IP system, https://www.kagg le.com/imenbenyahia/clearwatervnf-virtual-ip-multimedia-ip-system.} as the experimental dataset for performance evaluation. The entire dataset consists of 6 independent subsets of resource metrics data belonging to 6 different VMs. Each subset contains more than 170 thousand records where each record is described by 26 metric features (1st-4th are CPU-based, 5th-12th are disk-based, 13th-19th are memory-based, and 20th-26th are network-based), which are specified in Table 1. We partition each subset into three disjoint datasets: training dataset for model training, validation dataset for \emph{threshold} setting, and test dataset for performance evaluation.

\begin{table*}[htbp]
\centering
\caption{Metric Features in VNFDataset}
\renewcommand\arraystretch{1.32}
\scalebox{0.8}{\begin{tabular}{c c c c}
\hline
\hline
Feature types&Feature numbers&Detailed description\\
\hline
CPU-based & 4 &Including CPU idle percent, wait percent, system percent, stolen percent, etc.\\
Disk-based & 8 &Including disk used percent, read/write requests, read/write frequencies (times/s), read/write rates (kbytes/s), etc. \\
Memory-based & 7 &Including average loads, usable percent, usable size (MB), free size (MB), total size (MB), etc.\\
Network-based & 7 &Including in/out size (bytes/s), in/out errors (/s), in/out packets (/s), dropped packets (/s), etc.\\
\hline
\end{tabular}}
\end{table*}

2) \emph{Evaluation metrics}: To evaluate the performance of the proposed FL-based multi-discriminator BiWGAN-GP algorithm and the contrast algorithms on anomaly detection, the following metrics are considered as the comparison indicators, which can be categorized into two aspects: (1) Efficiency in terms of the communication, computation and memory cost on VM monitors, network slice managers and network controller in the training process; (2) Effectiveness in terms of Precision, Recall, F1-socre and Accuracy, which are formulated based on four fundamental indicators: TN (true negative) represents the number of correctly identified normal behaviors, FN (false negative) indicates the number of incorrectly identified abnormal behaviors as normal ones, FP (false positive) represents the number of incorrectly identified normal behaviors as abnormal ones, and TP (true positive) indicates the number of correctly identified abnormal behaviors.

\emph{Precision} refers to the ratio of correctly identified abnormal behaviors to the total identified abnormal ones, calculated by

\begin{equation}
Precision = \frac{{TP}}{{TP + FP}}.
\end{equation}

\emph{Recall} refers to the ratio of correctly identified abnormal behaviors to the total actual abnormal ones, calculated by
\begin{equation}
Recall = \frac{{TP}}{{TP + FN}}.
\end{equation}

\emph{F1-score} is the weighted average of the precision and recall, calculated by
\begin{equation}
F1 \textrm{-} score = 2 \times \frac{{precison \times recall}}{{precison + recall}}.
\end{equation}

Accuracy refers to the ratio of correctly identified behaviors to the total ones, calculated by
\begin{equation}
Accuracy = \frac{{TP + TN}}
{{TP + TN + FP + FN}}.
\end{equation}

\subsection{Efficiency}
\emph{Computational complexity}: We analyze the computational complexities of the proposed FL-based multi-discriminator BiWGAN-GP algorithm from three different sides, including VM monitors, network slice managers and the network controller, and compare them with the standalone BiWGAN-GP algorithm (only runs in VM monitors), multi-discriminator BiWGAN-GP algorithm (runs in VM monitors and network slice managers), and centralized BiWGAN-GP algorithm (only runs in the network controller), respectively. In the proposed algorithm, the computational complexity on VM monitors mainly relies on the architecture of discriminator $D$, and the computational complexities on network slice managers and the network controller mainly rely on the architectures of encoder $E$ and generator $G$. As $D$, $E$ and $G$ are all built using neural networks, their computational complexities are hard to accurately describe due to many free variables \cite{9076796}. For simplicity, we assume that the number of floating point operations to process one-batch data in $D$, $E$ and $G$ is directly proportional to the cardinality of model parameters $|\theta _D|$,  $|\theta _E|$ and $|\theta _G|$, respectively. Besides, the number of floating point operations to update parameters of  $D$, $E$ and $G$ is also assumed to be directly proportional to $|\theta _D|$,  $|\theta _E|$ and $|\theta _G|$, respectively. Therefore, the detailed computational complexities on VM monitors, network slice managers and the network controller are analyzed as follows:
\begin{itemize}
\item \emph{VM monitors}. On each VM monitor, the complexities of computing error feedbacks and updating model parameters of $D$ are $O(4M|\theta _D |)$ and $O(4KM|\theta _D |)$ per training iteration. Therefore, the total computational complexity on each VM monitor is $O(4I(1 + K)M|\theta _D |)$, which is much less than $O(4IKM|\theta _D |+ 2IM(|\theta _E | + |\theta _G |))$ of the standalone BiWGAN-GP algorithm on each VM monitor. As the multi-discriminator BiWGAN-GP algorithm is part of the proposed algorithm, they have the same computational complexity on VM monitors.
\item \emph{Network slice managers}. On each network slice manager, the complexity of encoding and generating one-batch data using $E$ and $G$ is $O(MN(|\theta _E | + |\theta _G |))$, and the complexity of updating model parameters of $E$ and $G$ is also $O(MN(|\theta _E | + |\theta _G |))$. Therefore, the total computational complexity on each network slice manager is $O(2IMN(|\theta _E | + |\theta _G |))$. Similarly, the proposed algorithm and the multi-discriminator BiWGAN-GP algorithm have the same computational complexity on network slice managers.
\item \emph{Network controller}. On the network controller, the proposed algorithm only needs to aggregate model parameters of $E$ and $G$ from all network slice managers every $L$ training iterations. Hence the computational complexity on the network controller is $O(S(|\theta _E | + |\theta _G |)I/L)$, which is much less than $O(2SIMN(2K|\theta _D |{\text{ + }}|\theta _E | + |\theta _G |))$ of the centralized BiWGAN-GP algorithm on the network controller.
\end{itemize}

\begin{figure*}[htbp]
\centering
\includegraphics[width=4.6in]{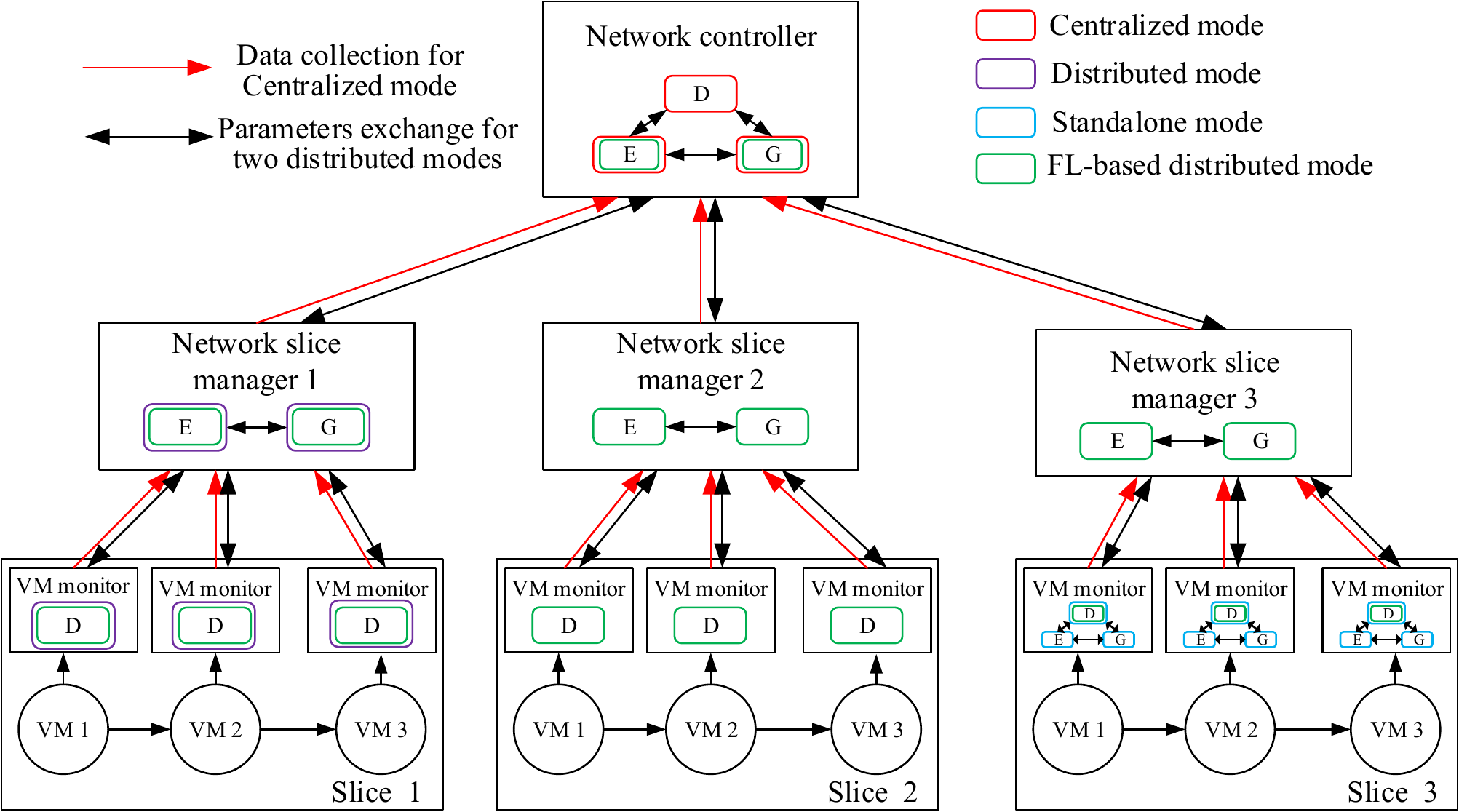}
\caption{The diagram of different training modes.}
\end{figure*}

\begin{figure*}[htbp]
    \centering
    \subfigure[]{\label{Fig:R1}
    \includegraphics[width=1.75in]{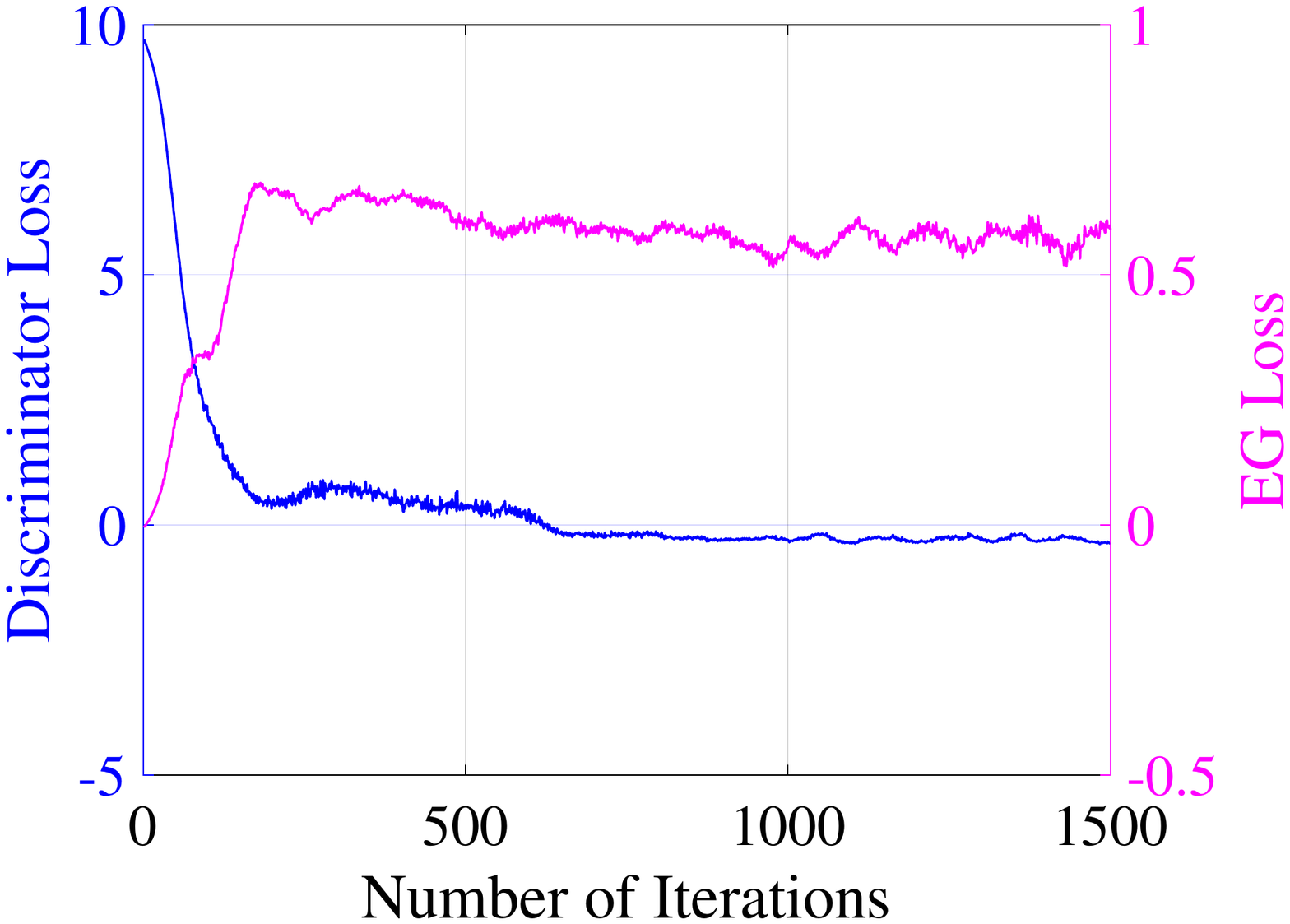}}
    \subfigure[]{\label{Fig:R2}
    \includegraphics[width=1.75in]{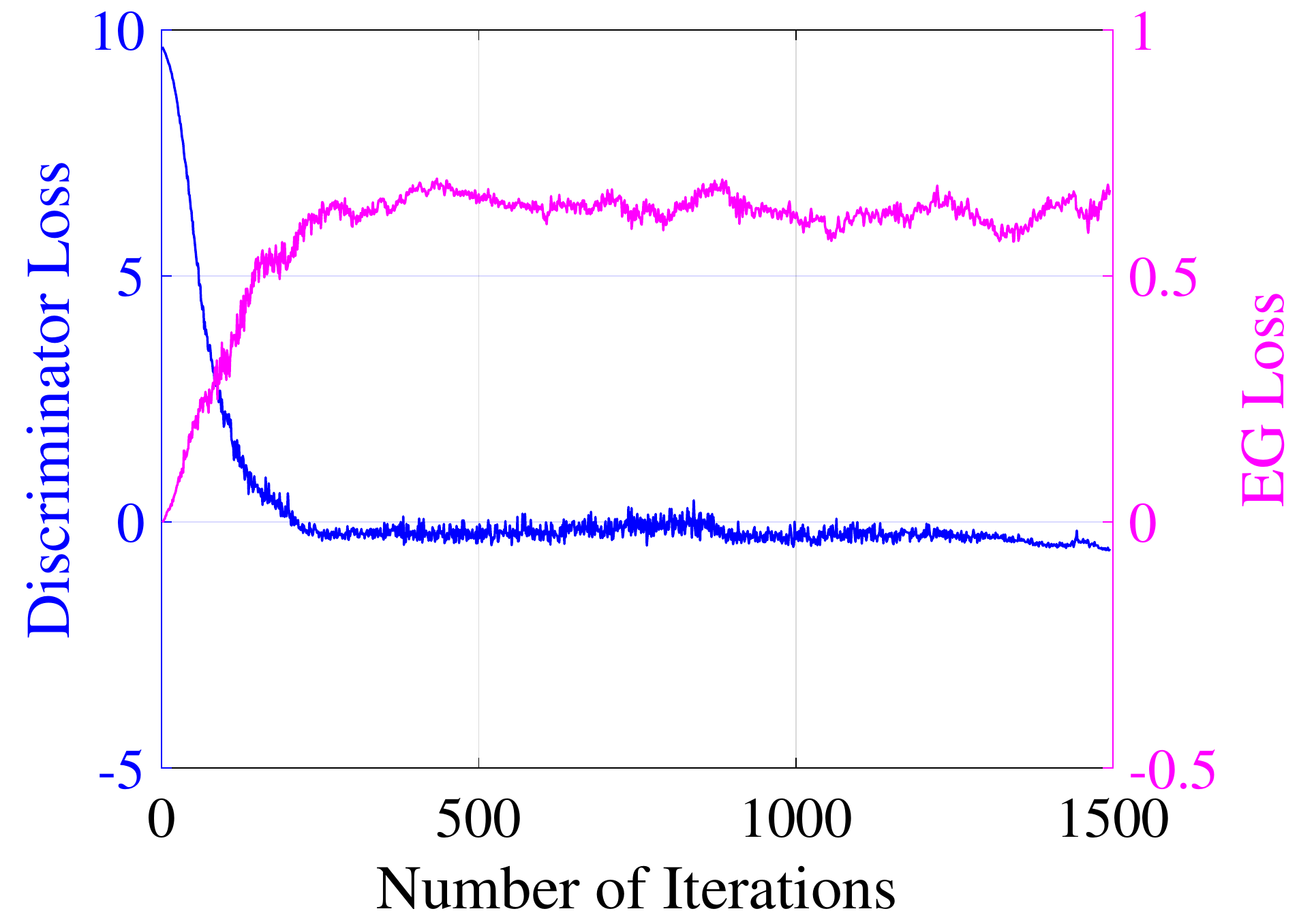}}
    \subfigure[]{\label{Fig:R3}
    \includegraphics[width=1.75in]{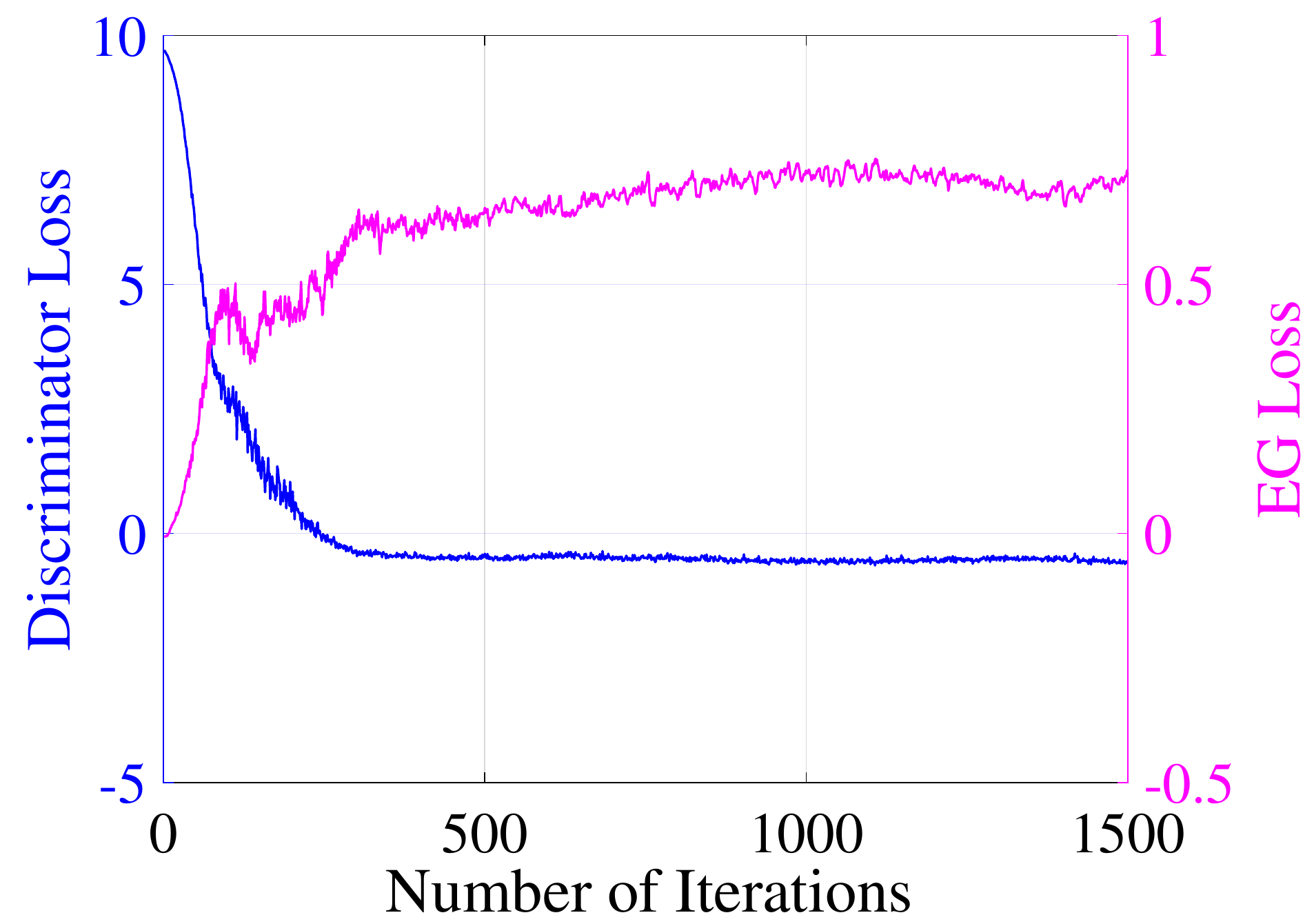}}
    \subfigure[]{\label{Fig:R4}
    \includegraphics[width=1.75in]{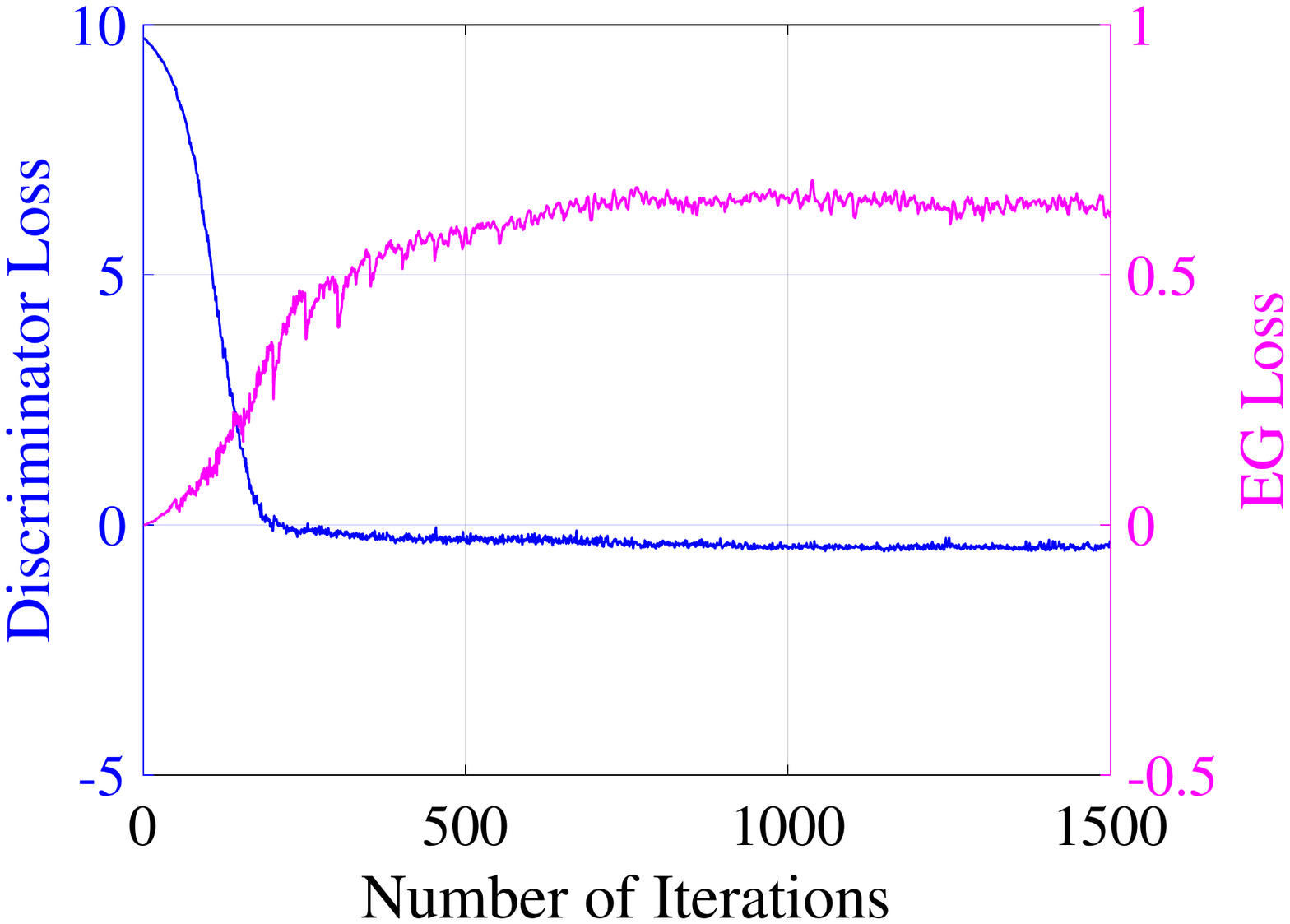}}
    \caption{Training processes of BiWGAN-GP model in different training modes. (a) Centralized training mode; (b) Standalone training mode; (c) Multi-discriminator BiWGAN-GP model in distributed training mode; (d) Multi-discriminator BiWGAN-GP model in FL-based distributed training mode.}
\end{figure*}

To validate the convergence of BiWGAN-GP model in different training modes, including centralized, standalone, distributed, and the proposed FL-based distributed modes as shown in Fig. 5, the averaged values of discriminator loss and encoder-generator (EG) loss in each iteration are recorded and depicted in Fig. 6. In centralized training mode, the network controller is responsible for collecting metrics data of all VMs and trains the centralized BiWGAN-GP model for anomaly detection. In standalone training mode, the BiWGAN-GP model is trained inside each VM monitor without any data or parameters exchange. In distributed training mode, the multi-discriminator BiWGAN-GP algorithm presented in Algorithm 1 is used for anomaly detection, where the network slice manager is responsible for the training of generator and encoder and VM monitors are responsible for the training of discriminators. The proposed FL-based distributed training mode is the combination of multiple distributed modes, which is elaborated in Algorithm 2. Fig. 6(a), (b), (c) and (d) shows the training processes of BiWGAN-GP model in centralized, standalone, distributed and FL-based distributed modes, respectively. The discriminator loss and EG loss both converge to stable values after about 300 training iterations, which indicates that the generators, encoders and discriminators all can be successfully trained in all four training modes. By contrast, the proposed FL-based distributed training mode enables the discriminator loss and EG loss to converge to the most stable values and achieve the best convergence performance because of its collaboration feature.

Fig. 7(a) shows the computation cost of the above four training modes under different number of iterations. The computation cost is measured by the consumed time for training. By contrast, the centralized training mode needs the most training time because it requires the network controller to collect metrics data from all VM monitors for the centralized  processing. The standalone training mode needs the least training time because each VM monitor trains the BiWGAN-GP model individually using its own local metrics data. Besides, the distributed and FL-based distributed training modes for multi-discriminator BiWGAN-GP model are more efficient than the centralized mode, but slightly inferior to the standalone mode.

In standalone mode, there is no communication among VM monitors, network slice managers and the network controller. In centralized mode, VM monitors send their all metrics data to the corresponding network slice managers and assemble on the network controller at the beginning. As the distributed and FL-based distributed training modes decentralize the discriminator training tasks to VM monitors by leaving generator and encoder training tasks on network slice managers, the communication cost between network slice managers and VM monitors is reasonable compared with the centralized mode. Fig. 7(b) shows the communication cost between network slice managers and VM monitors with the variation of batch sizes. The larger batch size brings a higher communication cost due to the increased data exchanged between network slice managers and VM monitors. Fig. 7(c) shows the communication cost between network slice managers and the network controller when increasing the number of local iterations $L$ before aggregation. The communication cost decreases with the increase of $L$ because the bigger $L$ means less interactions between network slice managers and the network controller.

Fig. 7(d) shows the memory cost for the above four training modes as the size of total training dataset increases. It shows that the memory cost on the network controller under distributed and FL-based distributed training modes is much less than that under the centralized training mode. Besides, it also indicates that the memory cost on the VM monitors under distributed and FL-based distributed training modes is also slightly less than that under the standalone training mode.
\begin{figure*}[htbp]
    \centering
    \subfigure[]{\label{Fig:R1}
    \includegraphics[width=1.76in]{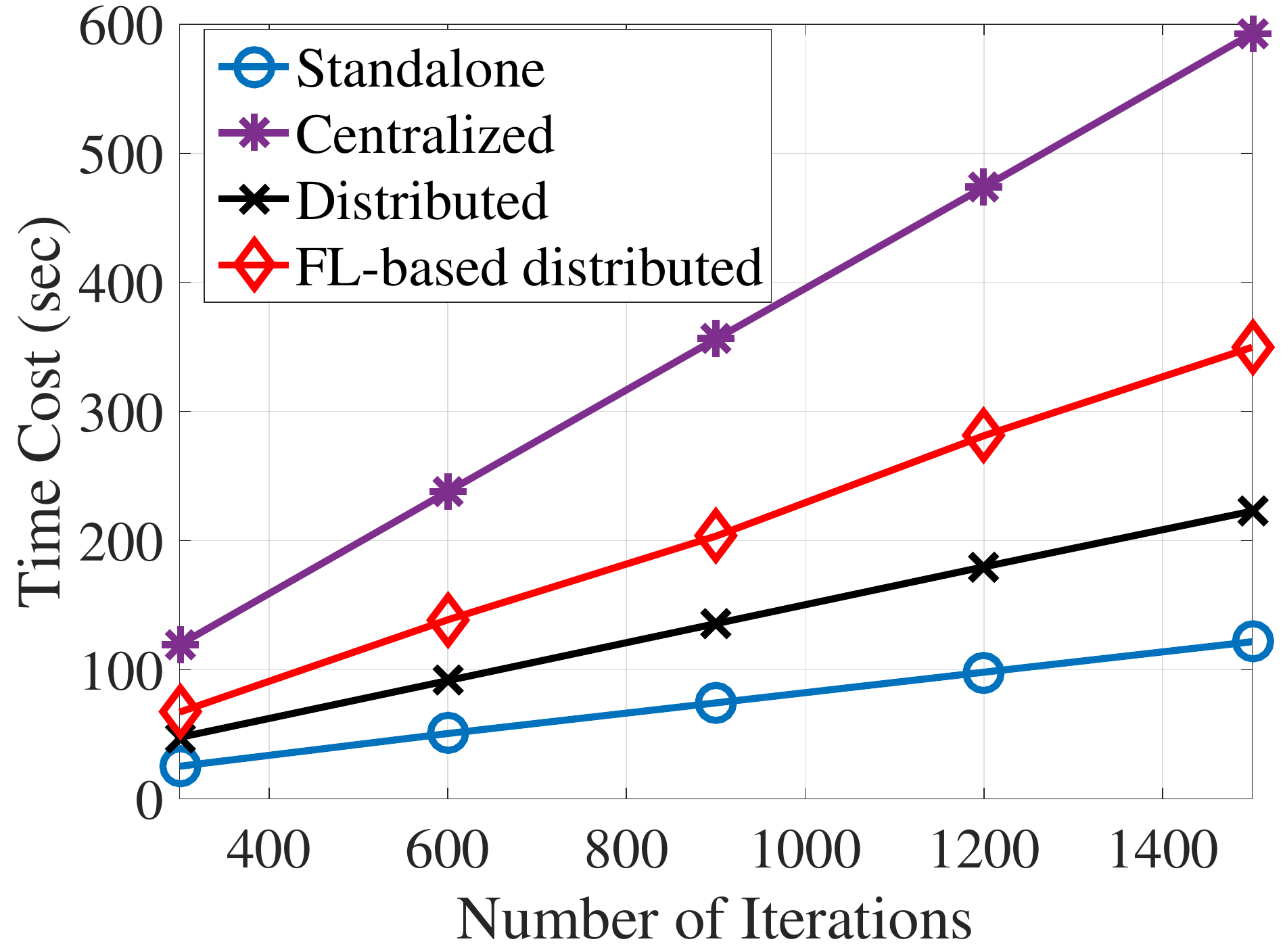}}
    \subfigure[]{\label{Fig:R2}
    \includegraphics[width=1.76in]{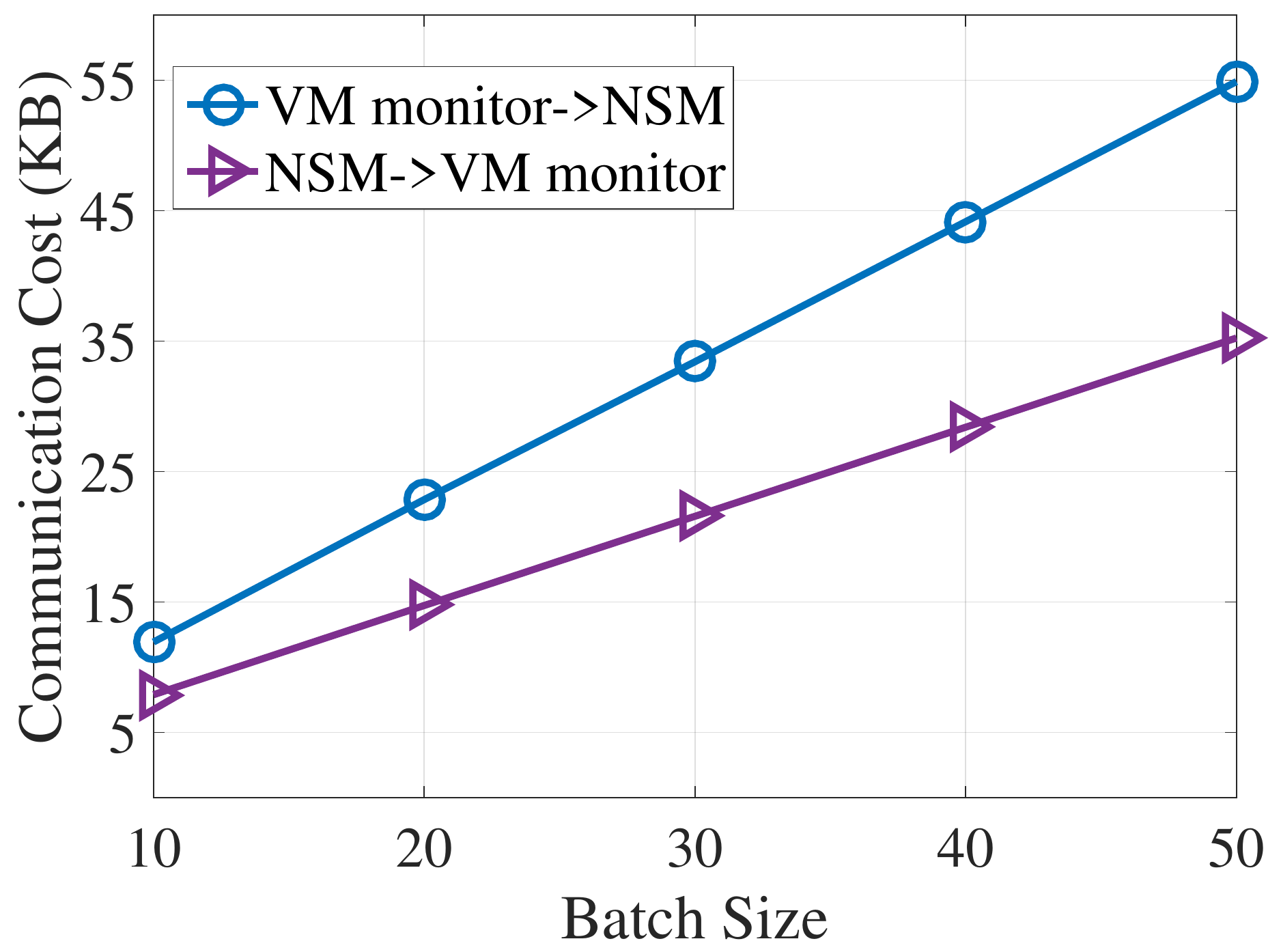}}
    \subfigure[]{\label{Fig:R3}
    \includegraphics[width=1.75in]{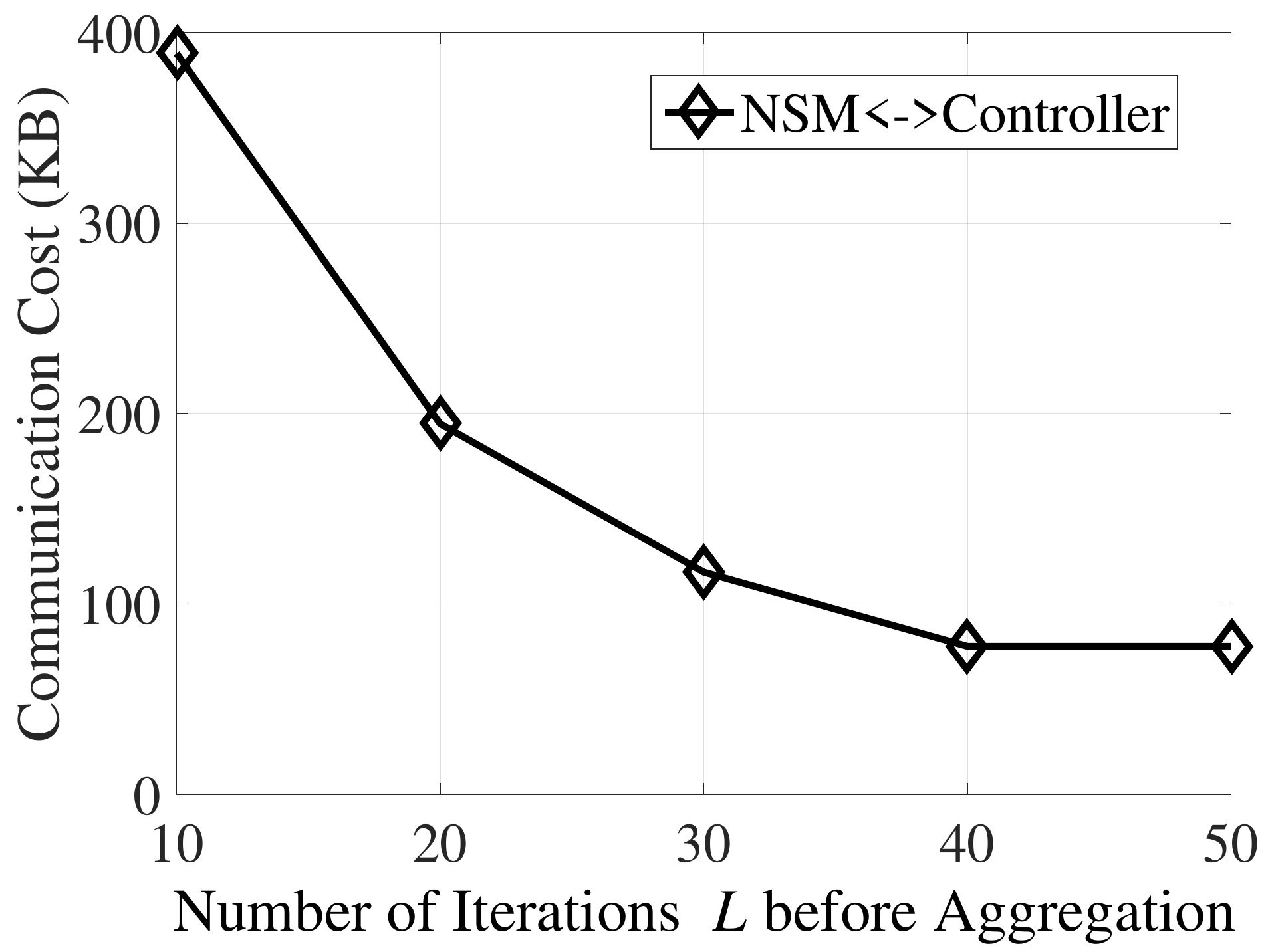}}
    \subfigure[]{\label{Fig:R4}
    \includegraphics[width=1.73in]{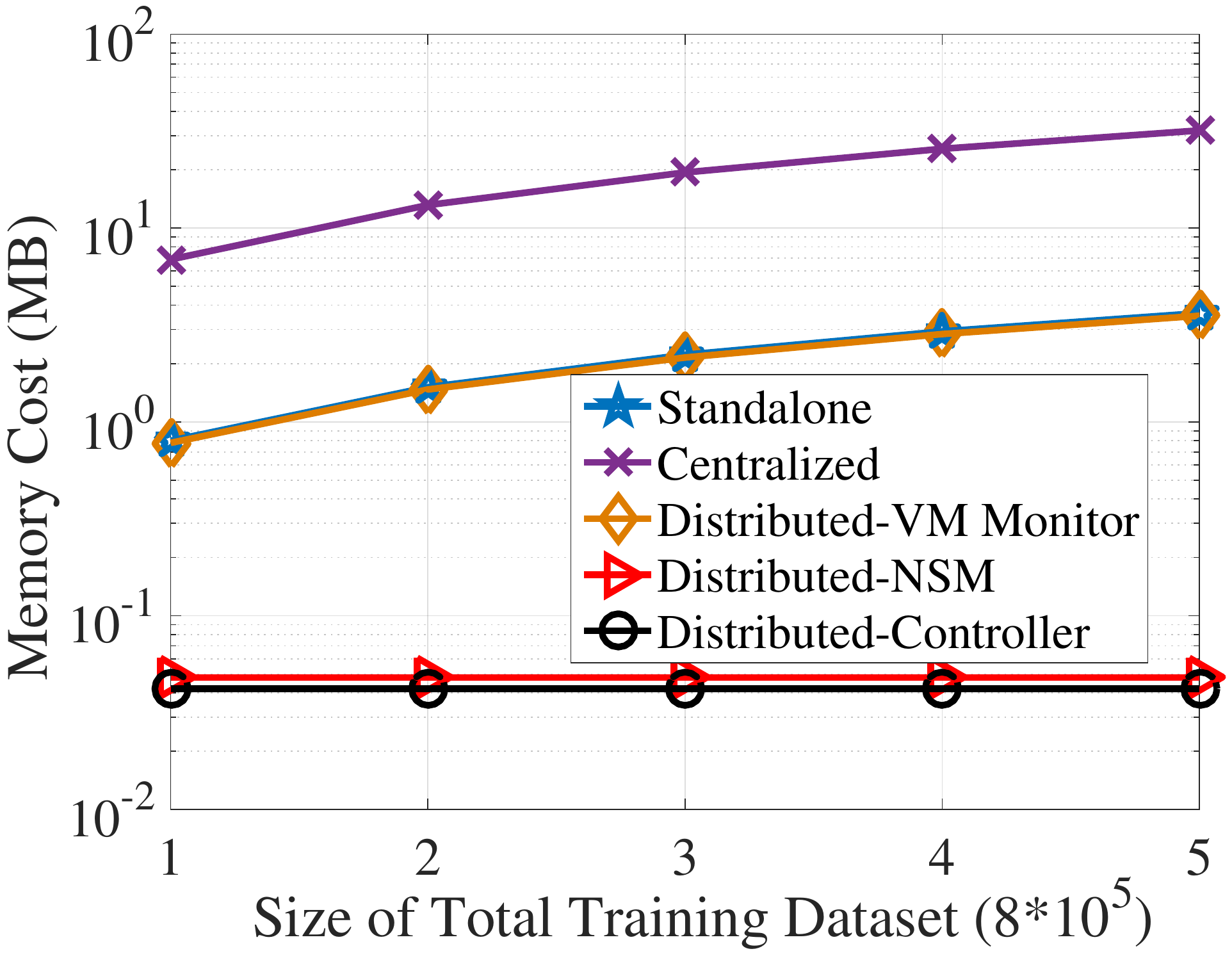}}
    \caption{Efficiency of different training modes. (a) Computation cost; (b) Communication cost between network slice managers (NSM) and VM monitors; (c) Communication cost between NSM and the network controller; (d) Memory cost.}
\end{figure*}

\subsection{Effectiveness}
 After the FL-based multi-discriminator BiWGAN-GP model is trained well, the abnormality of new metrics data can be measured through its anomaly score calculated by Eq. (28). If the anomaly score is larger than a preset $threshold$, new metrics data will be determined as abnormal. The value of $threshold$ will affect the detection performance of the FL-based multi-discriminator BiWGAN-GP model. The relationship between precision, recall and F1-score under different values of $threshold$ is presented in Fig. 8, where each point is obtained by setting a new value of $threshold$.
\begin{figure}[t!]
\centering
\includegraphics[width=2.45in]{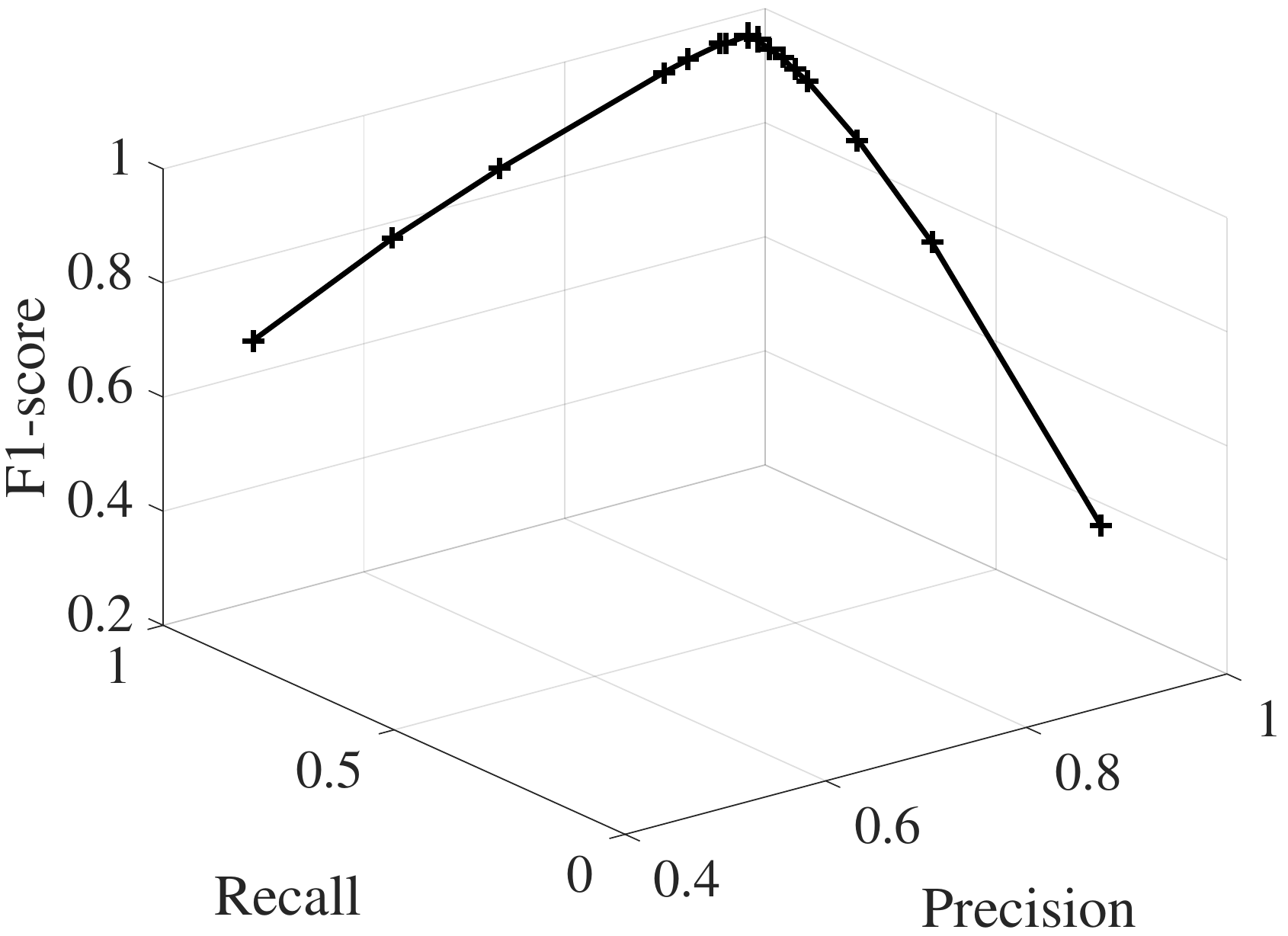}
\caption{The relationship between precision, recall and F1-score.}
\end{figure}

From Fig. 8, we can see that different $threshold$ values bring different detection performance, so we design a simple method to obtain an effective $threshold$  value using the validation dataset. To simulate anomalies in VMs, four error types are injected to the validation dataset, including endless loop in CPU, memory leak, Disk I/O fault and network congestion. Then, we calculate the average anomaly scores $A_{normal}$ and $A_{abnormal}$ of normal and abnormal metrics in the modified validation dataset. The $threshold$ is set as
\begin{equation}
threshold = \frac{{A_{normal}  + A_{abnormal} }}{2}.
\end{equation}

\begin{figure}[t!]
\centering
\includegraphics[width=2.7in]{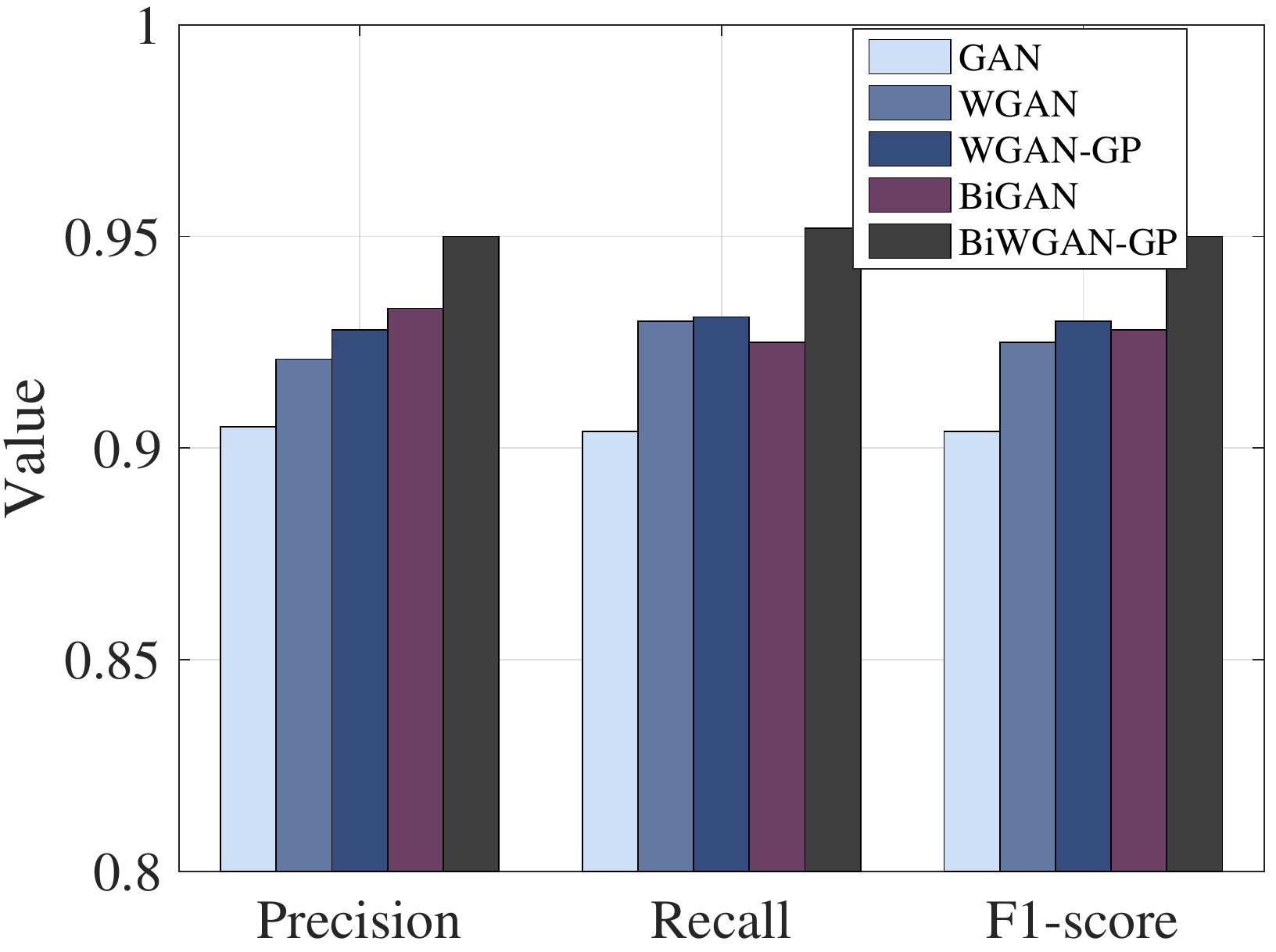}
\caption{The detection performance comparison between BiWGAN-GP and the state-of-the-art GAN algorithms.}
\end{figure}

BiWGAN-GP is the basic model used in the developed FL-based three-tier distributed anomaly detection framework. To validate the detection performance of the BiWGAN-GP algorithm and the state-of-the-art GAN algorithms in the presence of anomalies, four error types are injected to the test dataset randomly to simulate the anomalies in resource metrics data, including endless loop in CPU, memory leak, Disk I/O fault and network congestion. Fig. 9 shows the detection performance comparison between the BiWGAN-GP algorithm and the state-of-the-art GAN algorithms, including GAN [20], WGAN [21], WGAN-GP [23] and BiGAN [22]. For convenience, the standalone training mode is used for all the five algorithms in this simulation. Because the BiWGAN-GP algorithm combines both the strengths of BiGAN and WGAN-GP, it has obviously improved the detection performance compared to the state-of-the-art GAN algorithms. Specifically, compared to GAN, WGAN, WGAN-GP and BiGAN, the precision of BiWGAN-GP algorithm has increased by 4.5\%, 2.9\%, 2.2\%, and 1.8\%, respectively, the recall of BiWGAN-GP algorithm has increased by 4.8\%, 2.2\%, 2.1\%, and 2.7\%, respectively, and the F1-score of BiWGAN-GP algorithm has increased by 4.6\%, 2.5\%, 2.0\%, and 2.2\%, respectively.

\begin{figure}[t]
    \centering
    \subfigure[]{\label{Fig:R1}
    \includegraphics[width=2.45in]{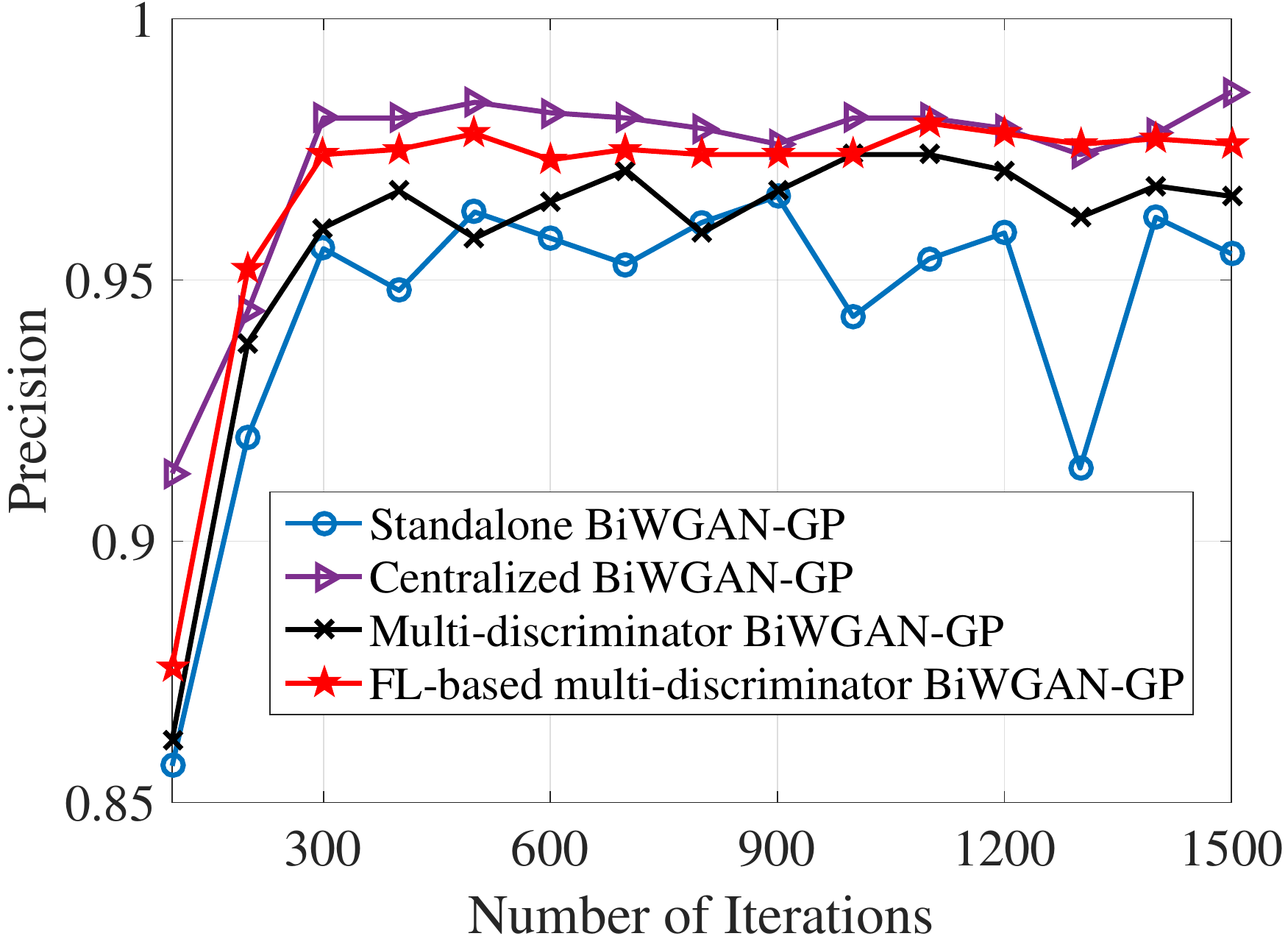}}
    \subfigure[]{\label{Fig:R2}
    \includegraphics[width=2.45in]{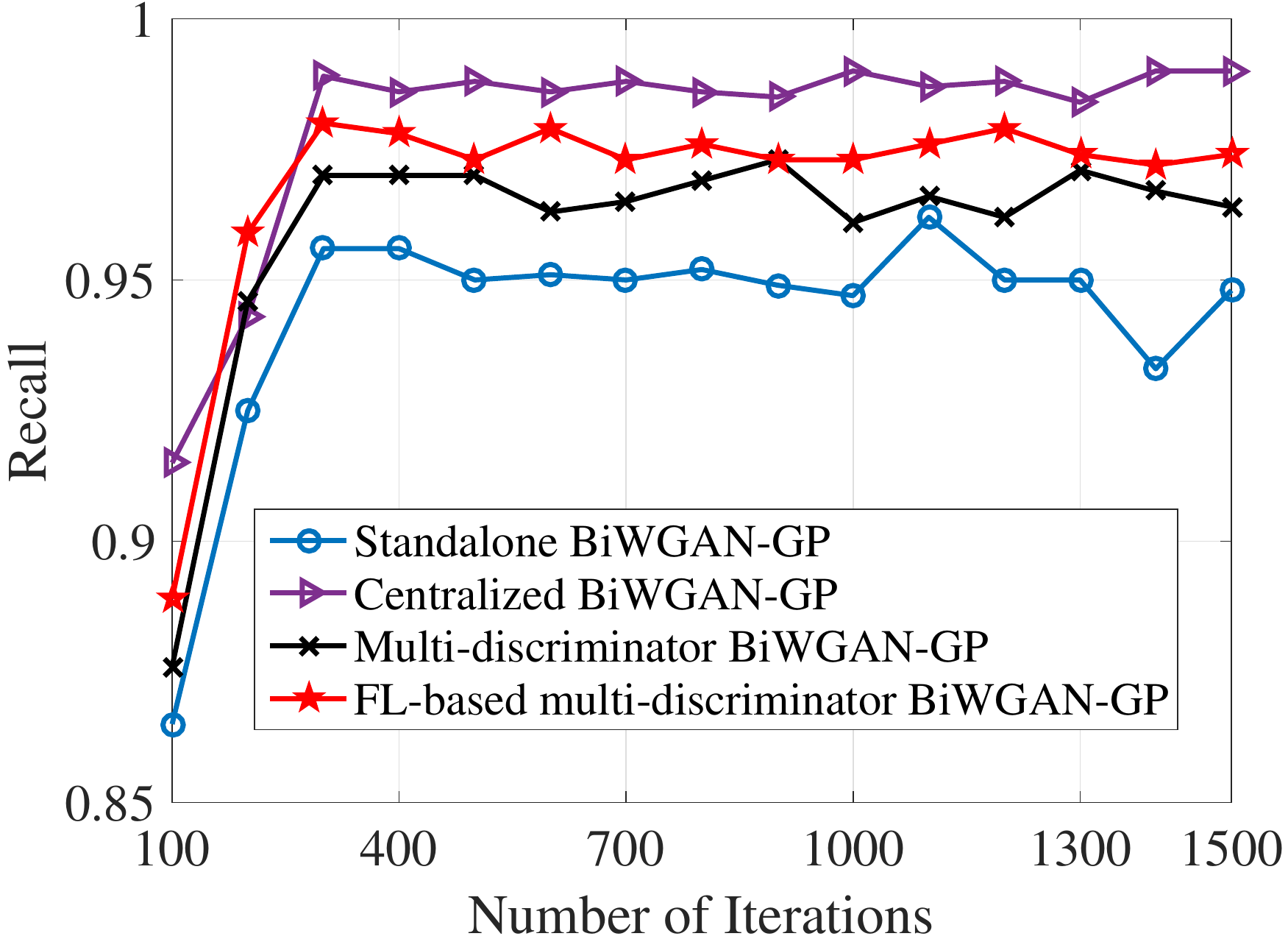}}
    \subfigure[]{\label{Fig:R3}
    \includegraphics[width=2.45in]{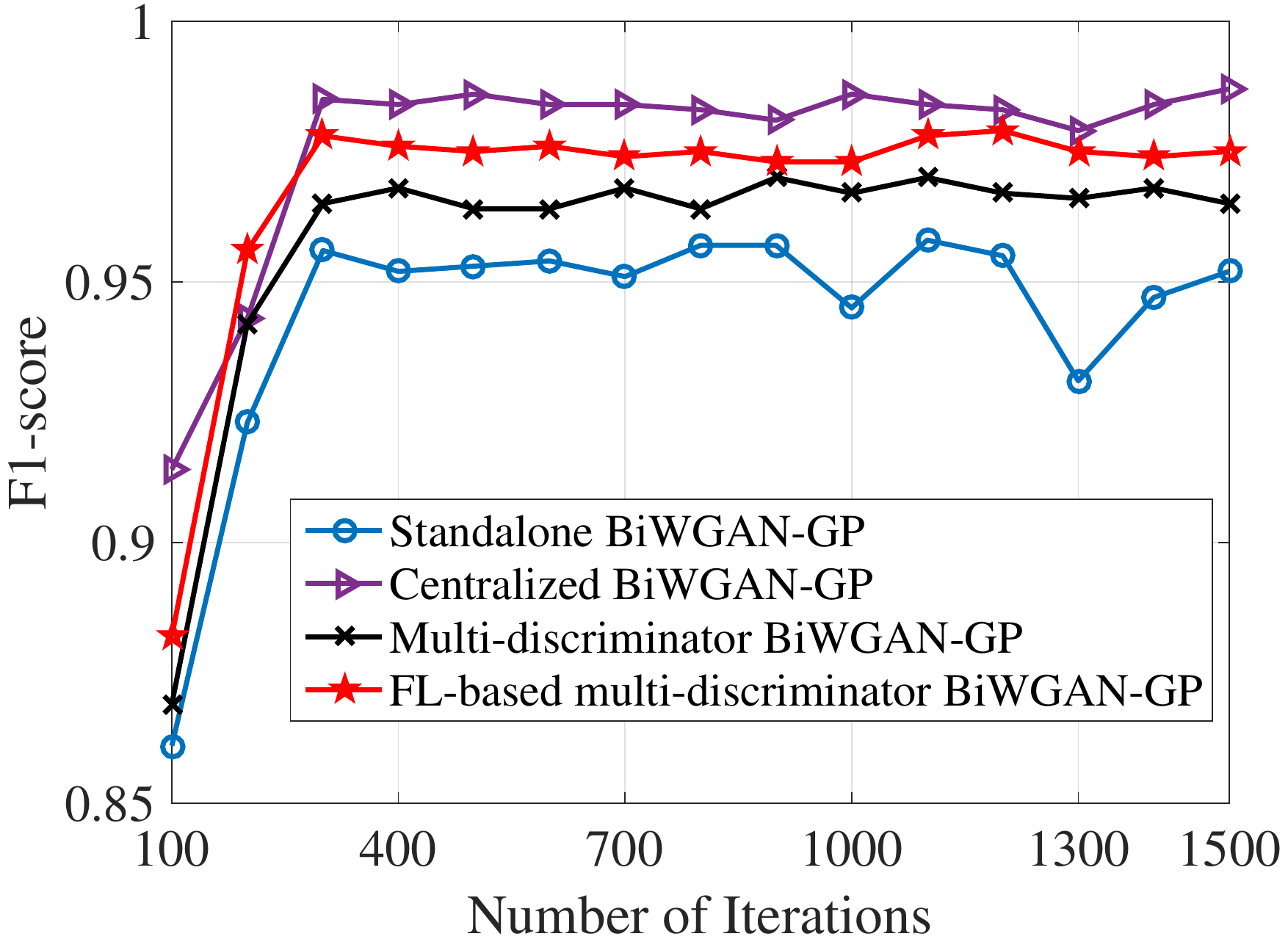}}
    \caption{The detection performance comparison among four algorithms. (a) Precision; (b) Recall; (c) F1-score.}
\end{figure}

Fig. 10 shows the detection performance of the centralized BiWGAN-GP, the standalone BiWGAN-GP, the multi-discriminator BiWGAN-GP and the FL-based multi-discriminator BiWGAN-GP algorithms. We use the precision, recall and F1-socre as the evaluation metrics. We can see that the values of three metrics using any algorithm are nearly stable after 300 iterations, which are consistent with the convergence trends of discriminator and EG losses presented in Fig. 6. It indicates that the four algorithms converge through enough iterations and can capture the distribution from metrics data of VMs. As shown in Fig. 10, the FL-based multi-discriminator BiWGAN-GP algorithm is superior to the standalone BiWGAN-GP and multi-discriminator BiWGAN-GP algorithms, and slightly inferior to the centralized BiWGAN-GP algorithm. Besides, according to the variation ranges of recall, precision and F1-score after 300 iterations, the detection performance of the FL-based multi-discriminator BiWGAN-GP algorithm is more stable than that of the standalone BiWGAN-GP and multi-discriminator BiWGAN-GP algorithms because of its collaboration feature.

The detection performance of four anomaly detection algorithms on four anomaly cases is summarized in Tables 2-5. Four anomaly detection algorithms are the centralized BiWGAN-GP, the standalone BiWGAN-GP, the multi-discriminator BiWGAN-GP and the FL-based multi-discriminator BiWGAN-GP algorithms. Four anomaly cases include endless loop in CPU, memory leak, Disk I/O fault and network congestion. By contrast, the performance of the FL-based multi-discriminator BiWGAN-GP algorithm is superior to the standalone BiWGAN-GP and multi-discriminator BiWGAN-GP algorithms, and slightly inferior to the centralized BiWGAN-GP algorithm on all four anomaly cases.

\begin{table}[t!]
\centering
\caption{Detection performance on endless loop in CPU}
\renewcommand\arraystretch{1.32}
\scalebox{0.8}{\begin{tabular}{c c c c c}
\hline
\hline
Algorithms & Accuracy & Precision & Recall & F1-score\\
\hline
Centralized & 0.9842 & 0.9833 & 0.9848 & 0.9841\\
Standalone & 0.9554 & 0.9549 & 0.9549 & 0.9549 \\
Multi-discriminator & 0.9651 & 0.9650 & 0.9645 & 0.9648\\
FL-based multi-discriminator & 0.9740 & 0.9710 & 0.9774 & 0.9742\\
\hline
\end{tabular}}
\end{table}

\begin{table}[t!]
\centering
\caption{Detection performance on memory leak}
\renewcommand\arraystretch{1.32}
\scalebox{0.8}{\begin{tabular}{c c c c c}
\hline
\hline
Algorithms & Accuracy & Precision & Recall & F1-score\\
\hline
Centralized & 0.9860 & 0.9769 & 0.9892 & 0.9853\\
Standalone & 0.9557 & 0.9563 & 0.9590 & 0.9560\\
Multi-discriminator & 0.9673 & 0.9664 & 0.9683 & 0.9674\\
FL-based multi-discriminator & 0.9743 & 0.9702 & 0.9783 & 0.9742\\
\hline
\end{tabular}}
\end{table}

\begin{table}[t!]
\centering
\caption{Detection performance on Disk I/O fault}
\renewcommand\arraystretch{1.32}
\scalebox{0.8}{\begin{tabular}{c c c c c}
\hline
\hline
Algorithms & Accuracy & Precision & Recall & F1-score\\
\hline
Centralized & 0.9874 & 0.9835 & 0.9916 & 0.9873\\
Standalone & 0.9559 & 0.9544 & 0.9570 & 0.9557\\
Multi-discriminator & 0.9667 & 0.9654 & 0.9672 & 0.9662\\
FL-based multi-discriminator & 0.9740 & 0.9718 & 0.9754 & 0.9736\\
\hline
\end{tabular}}
\end{table}

\begin{table}[t!]
\centering
\caption{Detection performance on network congestion}
\renewcommand\arraystretch{1.32}
\scalebox{0.8}{\begin{tabular}{c c c c c}
\hline
\hline
Algorithms & Accuracy & Precision & Recall & F1-score\\
\hline
Centralized & 0.9726 & 0.9713 & 0.9735 & 0.9726\\
Standalone & 0.9540 & 0.9501 & 0.9582 & 0.9542\\
Multi-discriminator & 0.9635 & 0.9609 & 0.9659 & 0.9634\\
FL-based multi-discriminator & 0.9740 & 0.9698 & 0.9754 & 0.9736\\
\hline
\end{tabular}}
\end{table}

\section{Related Work}
 A substantial part of the paper lies at the intersection of the three following research domains: 1) Monitoring framework for network virtualization environment; 2) GAN-based anomaly detection algorithm; 3) FL-based distributed learning paradigm.

\subsection{Monitoring framework for network virtualization environment}
 Network virtualization environment can reduce the deployment cost and simplify the life-cycle management of network functions, but it introduces new management challenges and issues \cite{8877749}. Therefore, building the effective monitoring framework for network virtualization environment is vital to ensure its high availability and reliability. To satisfy the security requirements of VMs, a comprehensive protection mechanism with the capacity of defense-in-depth for VMs was proposed in \cite{8359429}, where the anomaly detection was realized by building the normal traffic model and determining the abnormal behaviors according to their degree of deviation from the normal behaviors. Mishra \emph{et al}. \cite{8344510} designed a multi-level security architecture, namely VMGuard, for VM monitoring at the process level and system call level to detect known attacks and their variants. VMGuard applied the random forest classifier to produce a generic behavior for different categories of attacks in monitored VMs. As container-based virtualization has become a key solution in present virtualized networks, Zou \emph{et al}. \cite{8807263} developed an online container anomaly detection framework by monitoring and analyzing the resource-relevant metrics of containers with the optimized isolation forest algorithm. The considered metrics included CPU usage, memory usage, disk read/write speed and network speed. However, the monitoring framework designed in \cite{8359429,8344510,8807263} used centralized database to collect and process relevant data of VMs distributed in different regions of substrate networks, which will cause high communication and computation overhead when learning a global VM anomaly detection model. Our previous work \cite{9837465} designed a distributed monitoring mechanism from the perspective of physical nodes and links, not from the entire networks. In this paper, for each region of the entire networks, we deploy a VM monitor for a network slice to collect and store metrics data of VMs in the network slice as a local database. The distributed anomaly detection can be trained over datasets that are spread across the entire networks.

\subsection{GAN-based anomaly detection algorithm}
GAN is a promising generative model proposed by Goodfellow \emph{et al}. \cite{Goodfellow2014}, which is superior in capturing the distribution from high-dimensional complex real-world data. GAN consists of two models: the generator and the discriminator, which are usually built with neural networks. By training the generator and discriminator through the adversarial learning, the Nash equilibrium can be achieved theoretically \cite{8039016}, where the generator can generate fake samples sharing the same distribution with real ones that the discriminator is unable to distinguish between them. From this, GAN and its improved modifications have been widely used to implement anomaly detection in various fields, such as in distribution systems \cite{9108593,9139275}, in vehicular ad hoc networks \cite{9216536}, in high-speed railways \cite{8907821} and in movie reviews \cite{9076796}. The rationale behind the GAN-based anomaly detection is capturing the distribution from normal data and then finding an effective discriminant criterion to distinguish abnormal data from normal ones. In \cite{9108593}, BiWGAN was used as a feature extractor to map normal samples back to latent space and support vector data description was used to establish the discriminant criterion for nontechnical losses detection in power system. Using smart meters' data, a GAN was designed in \cite{9139275} to extract the normal temporal-spatial behavior of distribution systems. When abnormal behaviors occurred, the smart meters' data were expected to have anomaly scores out of the normal range. Shu \emph{et al}. \cite{9216536} proposed a distributed BiGAN framework to detect abnormal network behaviors for the entire vehicular ad hoc networks without exchanging data belonging to local sub-networks. In \cite{8907821}, a deep convolutional GAN was constructed to map the images of catenary support components into high-dimensional feature spaces, and an anomaly rating criterion was defined to detect abnormal images of high-speed railways. Gao \emph{et al}. \cite{9076796} developed a novel attention-driven conditional GAN to capture the correlations of movie reviews and identify spammed reviews. The effectiveness of GAN and its improved modifications in anomaly detection has been fully verified through extensive researches in these fields. In this paper, we design a new multi-discriminator BiWGAN-GP algorithm with multiple discriminators running locally in view of the distributed data features in virtualized network slicing environment.

\subsection{FL-based distributed learning paradigm}
FL is developed as a distributed leaning framework where training data are decentralized across learners, rather than being centralized \cite{8917592,9146846,9535454,9237167}. FL allows each learner to train a model from local data and send its local model to a center periodically. The center is responsible for model aggregation and sharing the global model with all learners. In \cite{8917592}, to model reliability in vehicular communications with low communication cost, FL was used to enable vehicular users to learn the tail distribution of network-wide queue lengths locally without sharing the actual queue data. As anomalies in edge devices seriously affect the manufacturing process in industrial IoT, a communication-efficient FL based deep anomaly detection framework was proposed in \cite{9146846} to enable distributed edge devices to collaboratively train a global detection model with improved generalization ability. Darwish \emph{et al}. \cite{9535454} extended FL to achieve the self-evolving network management for future intelligent vertical HetNet. In \cite{9237167}, based on the three-layer association relationship in radio access network slicing (device-base station-network slicing), a hybrid FL reinforcement learning framework was exploited to solve the device association problem while enforcing data security and device privacy, which included a horizontal aggregation on base stations for the same service types and a vertical aggregation on an encrypted center for different service types. Inspired by the existing researches, FL-based distributed learning paradigm can well adapt to the hierarchical control of virtualized network slicing, which facilitates the establishment of a global anomaly detection model without compromising the communication and computation efficiency.

\section{Conclusions}
 The novelty of the paper lies in the combination of multi-discriminator BiWGAN-GP and FL to detect anomalies in VMs in the context of virtualized network slicing environment. Within a network slice, the multi-discriminator BiWGAN-GP algorithm implements discriminators on VM monitors, and the generator and encoder on its network slice manager to monitor and analyze the resource metrics data of VMs in a distributed way. Then, the updated parameters of generators and encoders trained on network slice managers are sent to the network controller for aggregation using FL framework. We achieve a global VM anomaly detection model through the hierarchical cooperation among VM monitors, network slice managers and the network controller. We define an anomaly score using the reconstruction loss and discriminator loss of the multi-discriminator BiWGAN-GP algorithm, and then adopt the validation dataset to obtain the effective threshold for the anomaly score to distinguish between normal and abnormal behaviors. The experiment results on a real-world dataset validate the efficiency and effectiveness of the proposed FL-based multi-discriminator BiWGAN-GP algorithm on detecting typical anomalies of VMs in virtualized network slicing environment.


%


\ifCLASSOPTIONcaptionsoff
  \newpage
\fi



%
\bibliographystyle{IEEEtran}
 \bibliography{IEEEabrv,reference}

%




\end{document}